\documentclass{article}

\usepackage[latin1]{inputenc}
 \usepackage{amsmath,amssymb,amsfonts,amsthm,fullpage}

\newcommand{\reduceitem}{}
\newcommand{\reduceitemm}{}
\newcommand{\centermath}[1]{\smallskip\\\centerline{#1}\smallskip}

\newtheorem{theorem}{Theorem} 
\newtheorem{fact}{Fact} 
\newtheorem{definition}{Definition} 
\newtheorem{lemma}{Lemma} 
\newtheorem{proposition}{Proposition} 

\newsavebox{\fmbox}
\newenvironment{fmpage}[1]
     {\medskip\begin{lrbox}{\fmbox}\begin{minipage}{#1}}
     {\end{minipage}\end{lrbox}\fbox{\usebox{\fmbox}}\medskip}
\newcommand{\encadre}[1]{
\begin{center}
\begin{fmpage}{14cm}
#1
\end{fmpage}
\end{center}}

\newcommand{\e}{{\mathrm{e}}}
\newcommand{\complexi}{{\mathrm{i}}}

\newcommand{\complex}{\mathbb{C}}
\newcommand{\Span}{\mathsf{Span}}
\newcommand{\ket} [1] {\lvert #1 \rangle}
\newcommand{\bra} [1] {\langle #1 \rvert}
\newcommand{\braket}[2]{\langle #1\vert #2 \rangle}

\newcommand{\norm}[1]{\lVert #1 \rVert}
\newcommand{\size}[1]{\left\lvert #1 \right\rvert}
\newcommand{\abs}{\size}

\newcommand{\tensor}{\otimes}

\newcommand{\diag}{{\mathrm{diag}}}
\newcommand{\Order}{{\mathrm{O}}}
\newcommand{\order}{{\mathrm{o}}}
\newcommand{\adjoint}{\dagger}
\newcommand{\transpose}{\mathsf{T}}
\newcommand{\set}[1]{\left\{ #1 \right\}}
\newcommand{\ancilla}{\ket{\bar{0}}}

\newcommand{\suppress}[1]{}
\newcommand{\comment}[1]{}

\newcommand{\reflex}{\mathrm{ref}}

\newcommand{\id}{\mathrm{Id}}
\newcommand{\eps}{\varepsilon}
\newcommand{\pM}{p_M}
\newcommand{\aitch}{{\mathcal{H}}}
\newcommand{\ay}{{\mathcal{A}}}
\newcommand{\bee}{{\mathcal{B}}}
\newcommand{\ess}{{\mathcal{S}}}

\newcommand{\emm}{{\mathcal{M}}}

\newcommand{\emmt}{{\tilde{\mathcal{M}}}}
\newcommand{\kay}{{\mathcal{K}}}

\newcommand{\beq}{\begin{equation}}
\newcommand{\eeq}{\end{equation}}
\newcommand{\beqa}{\begin{eqnarray}}
\newcommand{\eeqa}{\end{eqnarray}}

\newcommand{\control}[1]{\mathrm{c}\mbox{-}#1}
\newcommand{\tmax}{{t_{\mathrm{max}}}}

\begin{document}

\title{Search via Quantum Walk\thanks{A preliminary version of this
    work appeared in {\em Proceedings of 39th ACM Symposium on Theory
      of Computing}, pages~575--584, 2007.}}

\author{
Fr\'ed\'eric Magniez\thanks{LIAFA, Univ. Paris 7, CNRS, F-75205 Paris, France.
       \texttt{magniez@liafa.jussieu.fr}}
\and
Ashwin Nayak\thanks{%
\suppress{
C\&O and IQC, U.\ Waterloo and Perimeter Institute;
200 University Ave.\ W., Waterloo, Ontario N2L 3G1, Canada.
       \texttt{anayak@uwaterloo.ca}
}
Department of Combinatorics and Optimization,
and Institute for Quantum Computing, University of Waterloo; and
Perimeter Institute for Theoretical Physics;
200 University Ave.\ W.,
Waterloo, ON, N2L 3G1, Canada.
Email: {\tt ashwin.nayak@uwaterloo.ca}.
}
\and
J\'er\'emie Roland\thanks{NEC Laboratories America;
      Princeton, NJ 08540, USA.
       \texttt{jroland@nec-labs.com}}
\and
Miklos Santha\thanks{LIAFA, Univ. Paris 7, CNRS, F-75205 Paris, France, and
Centre for Quantum Technologies,
National University of Singapore, Singapore 117543. \texttt{santha@lri.fr}}
}

\maketitle

\begin{abstract}
We propose a new method for designing quantum search algorithms for
finding a ``marked'' element in the state space of a classical Markov
chain. The algorithm is based on a quantum walk {\it {\`a} la\/}
Szegedy~(2004)
that is defined in terms of the Markov chain.
The main new idea is to apply quantum phase estimation to the quantum
walk in order to implement an approximate reflection operator. This
operator is then used in an amplitude amplification scheme. As a
result we considerably expand the scope of the previous approaches of
Ambainis~(2004) and Szegedy~(2004).
Our algorithm combines the benefits of these approaches in terms of being
able to find marked elements, incurring the smaller cost of the two,
and being applicable to a larger class of Markov chains.
In addition, it is conceptually simple and avoids some technical difficulties in the previous analyses of several algorithms based on quantum walk.
\end{abstract}


\section{Introduction}
\subsection{Background}

At an abstract level,
many search problems may be cast as the problem of
finding a ``marked'' element from a set~$X$ with~$n$ elements. Let~$M
\subseteq X$ be the set of
the so called
marked elements.
One approach to finding
an
element of~$M$,
if it is not empty, is to repeatedly sample from~$X$
uniformly until a marked element is picked. A more cost-effective
approach reuses resources expended in generating the first sample
(time, random bits, black-box queries, etc.) by simulating the steps
of a Markov chain with state space~$X$ to generate the next
sample. This approach often takes advantage of some structure present
in the ground set~$X$ and the Markov chain, and leads to a more
efficient algorithm. In this article, we study quantum analogues of
this randomized scheme.

There are several ways of defining quantum analogues of Markov chains,
including both discrete and continuous time versions (see, for
example,
Ref.~\cite{Szegedy04}
for a detailed introduction). We restrict
our attention to discrete time analogues.

Discrete time quantum walks emerged gradually in the field of quantum
algorithms.
On the line they are related to the quantum cellular automaton model
of Meyer~\cite{Meyer96}.  Watrous~\cite{Watrous01} introduced quantum
walks on regular graphs, and used them to show that randomized
logarithmic space is included in quantum logarithmic space.
Afterwards notions related to quantum walks, such as mixing time, and
deviation from
the
starting state, were studied for restricted graphs by
several
researchers~\cite{NayakV00,AmbainisBNVW01,AharonovAKV01,Richter07},
suggesting the possibility of speed-up of classical algorithms based
on random walk.

Shenvi, Kempe, and Whaley~\cite{ShenviKW03} 
pointed out the algorithmic potential of
quantum walks by designing a walk based
algorithm to emulate Grover Search~\cite{Grover96}. The first
algorithm using quantum walks that goes beyond the capability of
Grover Search is due to Ambainis~\cite{Ambainis04} for Element
Distinctness.  In his seminal paper he resolved the quantum query complexity of the
problem, settling a difficult question that had been open for several
years~\cite{BuhrmanDHHMSW05,AaronsonS04}.
Finally Szegedy~\cite{Szegedy04} developed a theory of quantum walk
based algorithms. He designed a quantum search algorithm based on any
symmetric, ergodic Markov chain that detects the presence of a marked
element. He defined a notion of quantum hitting time that is
quadratically smaller than the classical average hitting
time.
Since then, in the framework of Ambainis or Szegedy, many new
algorithms
with substantially better complexity emerged in a variety
of contexts~\cite{AmbainisKR05,MagniezSS07,BuhrmanS06,MagniezN07,DornT07}.

This work develops a new schema for quantum search algorithms, based
on any ergodic Markov chain.
We adapt the quantum analogue of classical Markov chains due to Szegedy 
to possibly non-symmetric Markov chains, but use it more in the style 
of the Ambainis algorithm.
Departing from the two algorithms,
however, we use quantum walks only indirectly. In conjunction with the
well known phase estimation algorithm~\cite{Kitaev95,KitaevSV02,CleveEMM98}, the
quantum walk helps us implement an approximate reflection
operator. This operator may then be used within amplitude
amplification algorithms~\cite{Grover96,BrassardHMT02,HoyerMW03} for
search. As a result, our work generalizes previous ones by extending
the class of possible Markov chains, and improving the complexity in
terms of its relation with the eigenvalue or singular value gap of the
related Markov chain.  In addition, our approach is conceptually
simple, avoids several technical difficulties in the analysis of the
earlier approaches, and leads to improvements in various aspects of
the algorithms.

\subsection{Two subtly different search algorithms}
\label{sec-classical-search}
We identify a Markov chain over state space~$X$ with its transition
matrix $P = (p_{xy})_{x,y\in X}$,
where $p_{xy}$ is the probability of 
transition from $x$ to $y$.
A chain is {\em irreducible} if every state is 
reachable
from every other state, and an irreducible chain is {\em ergodic} if
it is also aperiodic
(equivalently, its reachability graph is non-bipartite).
The eigenvalues of a Markov chain are at most 1 in magnitude.  By the
Perron-Frobenius theorem, an irreducible chain has a unique stationary
distribution~$\pi=(\pi_x)$, that is, a unique left eigenvector~$\pi$
with eigenvalue~$1$ and positive coordinates summing up to~$1$. If the
chain is ergodic, the eigenvalue~$1$ is the only eigenvalue of $P$
with magnitude $1$.  We denote by $\delta = \delta(P)$ the {\em
  eigenvalue gap\/} of $P$, that is $1- \lambda$, where
$\lambda = \lambda(P) = \max_{\nu \in \Lambda} \size{\nu}$,
where~$\Lambda$ is the set of eigenvalues of $P$ different from $1$.
The {\em
  time-reversed Markov chain}
$P^* = (p^*_{xy})$ of $P$ is defined by the equations $\pi_x p_{xy} =
\pi_y p^*_{yx}$.  The chain
$P$ is said to be {\em reversible\/} if $P^* = P$.  The Markov chain
$P$ is {\em symmetric} if 
$P = P^\transpose$ where $P^\transpose$ denotes the matrix transpose
of $P$.  The stationary distribution of any symmetric chain is the
uniform distribution.

The optimal quantum algorithm for Element Distinctness discovered by
Ambainis~\cite{Ambainis04} recasts the problem in terms of search for
a marked state in a Johnson graph defined by the problem instance. The
algorithm may be viewed as a quantum analogue of the following search
process, where $P$ is 
a Markov chain defined on state space~$X$.

\encadre{ {\bf Search Algorithm~1}
\mbox{ } 
\begin{enumerate}\reduceitem

\item Initialize~$x$ to a state sampled from a probability
distribution~$s$ over~$X$.

\item Repeat for~$t_2$ steps
\begin{enumerate}\reduceitemm

\item\label{item-check} If the state~$y$ reached in the previous step
  is marked,
  then stop and output~$y$.

\item Else, simulate~$t_1$ steps of the Markov chain~$P$ starting with
  the current state~$y$.

\end{enumerate}

\item If the algorithm has not terminated, stop, and output `no marked
element exists'.
\end{enumerate}
} 

The parameters~$t_1$ and~$t_2$ in the algorithm are determined by the
properties of the Markov chain and the marked subset~$M$. The idea
behind this algorithm is illustrated by considering an ergodic Markov
chain~$P$.
When~$t_1$ is large enough, the
state~$y$ in step~(\ref{item-check}) above is distributed
(approximately) according to the stationary distribution of~$P$. Then,
the outer loop represents sampling from the stationary distribution
until a marked element is found.
When~$t_2$ is chosen to be inversely
proportional to the probability that a state is marked according to
the stationary distribution, the algorithm 
succeeds with high probability.

The analysis of the Ambainis quantum algorithm depends heavily on the form of marked
states, and was presented for subsets~$M$ arising out
of~$k$-Collision, a generalization of Element Distinctness, with the
assumption of a unique collision.  Inspired by this algorithm,
Szegedy~\cite{Szegedy04} designed a quantum search algorithm 
with uniform initial distribution, based on
any symmetric, ergodic Markov chain. The Szegedy algorithm may be
viewed as a quantum analogue of a subtly different, but more natural,
classical process.

\encadre{ {\bf Search Algorithm~2}
\mbox{ } 

\begin{enumerate}\reduceitem

\item Initialize~$x$ to a state sampled from a probability
distribution~$s$ over~$X$.

\item Repeat for~$t$ steps
\begin{enumerate}\reduceitemm

\item If the state~$y$ reached in the previous step 
 is marked,
  then stop and output~$y$.

\item Else, simulate {\em one\/} step of the Markov chain~$P$ from the
  current state~$y$.

\end{enumerate}

\item If the algorithm has not terminated, stop, and output `no marked
element exists'.

\end{enumerate}
} 

The parameter~$t$  is also determined by the Markov
chain $P$, and the set~$M$ of marked states. This algorithm is a
greedy version of the first algorithm: a check is performed after
every step of the Markov chain to determine if a marked state has been
reached, irrespective of whether the Markov chain has mixed.  

Let us formally derive the complexity of the two algorithms to clarify
their differences.  
Assume that the
search algorithms maintain a data structure~$d$ that associates some
data~$d(x)$ with every state $x \in X$.  {From} $d(x)$, we would like
to determine if $x \in M$.  When operating with $d$, 
we distinguish three types of cost.
\begin{itemize}\reduceitem

\item[] \textbf{Set-up cost $\mathrm{S}$:}
The cost of sampling~$x \in X$ according to the initial distribution~$s$
and of constructing the data structure~$d(x)$ for the state~$x$.

\item[] \textbf{Update cost $\mathrm{U}$:}
The cost of simulating a transition from~$x$ to~$y$ for a state~$x \in X$ 
according to the Markov chain~$P$ and of updating $d(x)$ to~$d(y)$.

\item[] \textbf{Checking cost $\mathrm{C}$:} The cost of checking
  if~$x \in M$ using~$d(x)$.

\end{itemize}
These costs may be thought of as vectors 
listing all the measures of complexity of interest, such as query and
time complexity.  We may now state generic bounds on the efficiency of
the two search algorithms in terms of our cost parameters. Note that 
throughout this paper, we say that an event happens \emph{with high 
probability} if it happens with probability at least
some universal constant.
All the search algorithms (classical and quantum) we discuss have 
one-sided error. The algorithms may fail with some probability to report any
marked element even when they exist. This error probability may be driven down
to the desired level in the standard manner by sequential repetition of 
the algorithms.

\begin{proposition}
\label{thm-classical-search}
Let~$\delta > 0$ be the eigenvalue gap of an ergodic, symmetric Markov
chain~$P$ on a state space~$X$ of size~$n$, and let
$\frac{\size{M}}{\size{X}} \ge \eps > 0$ whenever~$M \subset X$ is
non-empty. For the uniform initial distribution~$s$,
\begin{enumerate}\reduceitemm

\item \textbf{Search Algorithm~1} determines if a marked element
exists and finds one such element with high probability 
if~$t_1 \in \Order ( \tfrac{1}{\delta} )$ and~$t_2 \in \Order(
\frac{1}{\eps} )$ are chosen to be suitably large.
The cost incurred is of order
$
\mathrm{S} + \frac{1}{\eps} \left(
\frac{1}{\delta}\mathrm{U}+\mathrm{C} \right).
$
\item \textbf{Search Algorithm~2} determines if a marked element
exists and finds one such element with high probability 
if~$t \in \Order( \frac{1}{\delta \eps} )$ is chosen to be suitably large.
The cost incurred is of order
$
\mathrm{S} + \frac{1}{\delta\eps} \left( \mathrm{U}+\mathrm{C} \right).
$

\end{enumerate}\reduceitemm
\end{proposition}

{
\begin{proof}
The stopping time of \textbf{Search Algorithm~2} is the average
hitting time of the set~$M$ for the Markov chain~$P$. We may therefore
take~$t$ to be a constant factor more than this hitting time.  As
mentioned before, this time is bounded above by the stopping time for
the first algorithm. Therefore part~2 of the proposition follows from
part~1.

In the first algorithm, we may take~$t_2$ to be proportional to the
average hitting time of the set~$M$ for the Markov
chain~$P^{t_1}$. 
The quantity~$\lambda(P)$ is bounded by~$1 - \delta$ by
hypothesis. The analogous quantity~$\lambda(P^{t_1})$ is therefore
bounded by~$(1-\delta)^{t_1} \le \e^{-\delta t_1}$.
Taking~$t_1 = 1/\delta$, we get a spectral gap~$\tilde{\delta}$ of at
least~$1/2$ for~$P^{t_1}$. We may now bound the average hitting time
of~$M$ for~$P^{t_1}$ by, for example, Equation~(15) in
Ref.~\cite{Szegedy04} and Lemma~1 in Ref.~\cite{BroderK89} (also
stated as Lemma~10 in Ref.~\cite{Szegedy04}). This bound evaluates
to~$\tfrac{1}{\eps \tilde{\delta}} \le \tfrac{2}{\eps}$. The
expression for the cost of the algorithm now follows.
\end{proof}
}
For special classes of graphs, for example for the 2-d toroidal grid,
the hitting time may be significantly smaller than the generic
bound~$t = \Order(1/\delta\eps)$ given in part~2 (see
Ref.~\cite[Page~11, Chapter~5]{AldousF}).

\subsection{Quantum analogues}
\label{sec-q-analogue}
As in the classical case, the quantum search algorithms look for a marked element in
a finite set $X$, where a data structure $d$ is maintained during the algorithm.
Let $X_d$ be the set of items along with their associated data, that is $X_d=\{(x,d(x)):x\in X\}$.
For 
convenience
we suppose that $\bar 0\in X$ and that $d(\bar 0)=\bar 0$.

The quantum walks due to Ambainis and Szegedy, as in our work, may be
thought of as walks on {\em edges\/} of the original Markov chain,
rather than its vertices. Thus, 
the associated
state space is a linear subspace of the vector space~$\aitch =
\complex^{X \times X}$, or~$\aitch_d = \complex^{X _d\times X_d} $
when we also include the data structure.  For the sake of elegance in
the mathematical analyses, our data structure keeps the data for both
vertices of an edge, whereas in previous works the data was kept
only for one of them.

There is a natural isomorphism $\ket{\psi}\mapsto \ket{\psi}_d$
between $\aitch$ and $\aitch_d$, where on basis states
$\ket{x}_d=\ket{x,d(x)}$. This isomorphism maps a unitary operation
$U$ on $\aitch$ into $U_d$ on $\aitch_d$ defined by
$U_d\ket{\psi}_d=(U\ket\psi)_d$.  Our walks are discussed in the space
$\aitch_d$ when, for implementation and cost considerations, it is
important to properly deal with the data structure.  However, for
convenience, we analyze the mathematical properties of the walks
without the data structure, in the space $\aitch$. This is 
justified by
the isomorphism between $\aitch_d$ and $\aitch$.

The initial state of the algorithm is explicitly related to the stationary
distribution $\pi$ of $P$.
At each step, the right end-point of
an edge~$(x,y)$ is ``mixed'' over the neighbors of~$x$, and then the
left end-point is mixed over the neighbors of the new right
end-point. 
We again distinguish three types of cost generalizing those of the
classical search.  They are of the same order as the corresponding
costs in the algorithms of Ambainis and Szegedy.
Some operations of the algorithms not entering into these costs are
not taken into account.
This is justified by the fact that in all quantum
search algorithms the overall complexity
is of the order of the
accounted part, which is expressed in terms of the costs below.

\begin{itemize}\reduceitem

\item[] \textbf{(Quantum) Set-up cost $\mathsf{S}$:} The cost
of
constructing the state $\sum_x \sqrt{\pi_x}\ket{x}_d\ancilla_d$
from $\ancilla_d\ancilla_d$.

\item[] \textbf{(Quantum) Update cost $\mathsf{U}$:} The cost
of realizing
any of the unitary transformations
\smallskip\\
\centerline{$\ket{x}_d \ancilla_d 
    \quad \mapsto \quad \ket{x}_d
                        \sum_y\sqrt{p_{xy}}\ket{y}_d,$}
\centerline{$\ancilla_d \ket{y}_d
    \quad \mapsto \quad 
                        \sum_x\sqrt{p^*_{yx}}\ket{x}_d\ket{y}_d,$}
\smallskip
and their inverses, where $P^*=(p^*_{xy})$ is the time-reversed Markov
chain defined in Section~\ref{sec-classical-search}.

\item[] \textbf{(Quantum) Checking cost $\mathsf{C}$:}
The cost of realizing the following conditional phase flip
\[
\ket{x}_d\ket{y}_d \quad \mapsto \quad
\left\{
\begin{array}{rl}
-\ket{x}_d\ket{y}_d & \textrm{if } x\in M, \\
\ket{x}_d\ket{y}_d  & \textrm{otherwise.}
\end{array}
\right.
\]
\end{itemize}

The quantum search algorithms 
due to
Ambainis and Szegedy give a quadratic speed
up in the times~$t_1, t_2$ and~$t$, with respect to the classical algorithms.
Let us recall that for integers $0 < r < m$ and~$0 < l < r$ the vertices
of the Johnson graph with parameters $m,r,l$ are the subsets of size
$r$ of a universe of size $m$, and there is an edge between two
vertices if the size of their intersection is~$l$.
The eigenvalue gap $\delta$ of the symmetric walk on the Johnson graph
with~$l = r-1$, and~$r < m/2$ is in $\Theta(1/r)$. If the set of
marked vertices consists of vertices that contain a fixed subset of
constant size~$k \leq r$, then their
fraction~$\eps$ is in~$\Omega(\frac{r^k}{m^k})$.

\begin{theorem}[Ambainis~\cite{Ambainis04}]
\label{thm-ambainis}
Let~$P$ be the random walk on the Johnson graph on~$r$-subsets of a
universe of size~$m$, where~$r = \order(m)$, and with intersection
size~$r-1$. Let~$M$ be either empty, or the class of all~$r$-subsets
that contain a fixed subset of constant size~$k \leq r$. 
Then, there is a quantum algorithm that with high probability, determines if $M$ is empty or finds the~$k$-subset, with cost of order $ \mathsf{S} +
\frac{1}{\sqrt{\eps}} ( \frac{1}{\sqrt{\delta}}
\mathsf{U}+\mathsf{C}).  $
\end{theorem}

\begin{theorem}[Szegedy~\cite{Szegedy04}] 
\label{thm-szegedy}
Let~$\delta > 0$ be the eigenvalue gap of an ergodic, symmetric Markov
chain~$P$, and let $\frac{\size{M}}{\size{X}} \ge \eps > 0$
whenever~$M$ is non-empty.  
There exists a quantum
algorithm that determines, with high probability, if $M$ is non-empty with cost of order
$
\mathsf{S} + \frac{1}{\sqrt{\delta\eps}} (\mathsf{U}+\mathsf{C}).
$
\end{theorem}

If the checking cost~$\mathsf{C}$ is substantially greater than
that of performing one step of the walk, an algorithm with the cost
structure of the Ambainis algorithm would be more efficient.
Moreover, the algorithm would {\em find\/} a marked element if one
exists. These advantages are illustrated by the algorithm for Triangle
Finding~\cite{MagniezSS07}. This algorithm uses two quantum walks {\em
  \`a la\/} Ambainis recursively; the Szegedy framework seems to give
a less efficient algorithm.  Nonetheless, the Szegedy approach has
other advantages---it applies to a wider class of Markov chains and
for arbitrary sets of marked states.  Moreover, the
quantity~$1/\sqrt{\delta\epsilon}$ in Theorem~\ref{thm-szegedy} may be
replaced by the square-root of the classical hitting
time~\cite{Szegedy04}.  These features make it more suitable for
applications such as the near-optimal algorithm for Group
Commutativity~\cite{MagniezN07} which has no equivalent using the
Ambainis approach.

\subsection{Contribution, relation with prior work,
and organization}
\label{sec-contribution}

We present an algorithm that is
a quantum analogue of {\bf Search Algorithm 1} and
works for any ergodic Markov chain. It is most easily described for
{\em reversible\/} Markov chains.

\begin{theorem}
\label{thm-without-log-factor}
Let~$\delta > 0$ be the eigenvalue gap of a reversible, ergodic Markov
chain $P$, and let $\eps > 0$ be a lower bound on the probability that
an element chosen from the stationary distribution of $P$ is marked
whenever~$M$ is non-empty. Then, there is a quantum algorithm that
with high probability, determines if 
$M$ is empty or finds an element of $M$, with cost of
order
$
\mathsf{S} + \frac{1}{\sqrt{\eps}} ( \frac{1}{\sqrt{\delta}}
\mathsf{U}+\mathsf{C}).
$
\end{theorem}
This
algorithm considerably expands the scope of the approaches
embodied in Theorems~\ref{thm-ambainis} and~\ref{thm-szegedy} above.
It combines the benefits of 
the two
approaches in terms of being able to
find marked elements, incurring the smaller cost of the two, and being
applicable to a larger class of Markov chains. In addition, it is
conceptually simple, avoids several technical difficulties in the
analysis of the earlier approaches, and leads to improvements in
various aspects of algorithms for Element Distinctness, Matrix Product
Verification, Triangle Finding, and Group Commutativity.  Namely, we
give a single-shot method for any algorithm {\em \`a la\/} Ambainis in
presence of multiple solutions,
without the need for a reduction to special cases such as that of a
unique solution. This applies to Element Distinctness and Triangle
Finding. For Element Distinctness, Matrix Product Verification, and
Group Commutativity, where an algorithm {\em \`a la\/} Szegedy only
detects the existence of a solution, we find one with the same time
and query complexity.  Finally, we improve the
query complexity of the best
previously
known algorithm for Triangle Finding
by a $\mathrm{polylog}(n)$ factor.

In Section~\ref{sec-quantum-mc}, we describe
a quantum analogue of a Markov chain based on the work of
Szegedy~\cite{Szegedy04} who defined such a
quantum process $W(P,Q)$ for
a classical bipartite walk $(P,Q)$. By letting $Q=P$, he related the
spectrum of the quantum walk $W(P)$ to that of~$P$ for symmetric
Markov chains. 
Using an absorbing version of~$P$ as in \textbf{Search
Algorithm~2}, he designed a quantum analogue of this classical scheme.
Even when $P$ is not symmetric, letting $Q=P^{*}$, the time-reversed Markov
chain corresponding to~$P$,
leads to a
natural connection between
$P$ and $W(P)$.  If $P$ is
reversible, then the eigenvalues of $W(P)$ are closely related to
those of $P$, as in the symmetric case.
For an arbitrary, possibly non-reversible, ergodic Markov chain, this
connection relates the eigenvalues of~$W(P)$ to the singular values of
a ``discriminant'' matrix~$D(P)$ associated with~$P$.

In Section~\ref{sec-walk-search}, we use the quantum walk $W(P)$
associated 
with the unperturbed walk $P$ in a completely different way, more in the
style of the Ambainis approach.
Ambainis directly uses a power of $W(P)$ to replace the ``diffusion'' operator
in the Grover search algorithm.  The beauty of this step, and the 
difficulty of proving its correctness, lies in the fact that even if no power
of $W(P)$ closely approximates the diffusion operator, some powers 
have sufficient properties to mimic its essential features (see Lemma~3
in Ref.~\cite{Ambainis04}).
While this lemma is sufficient to prove Theorem~\ref{thm-ambainis}, it alone
is not powerful enough to imply Theorem~\ref{thm-without-log-factor}. 
The spectral gap of classical Markov chains and that of some special cases
of quantum walks (such as the quantum walk on Johnson graphs proposed by
Ambainis) may be amplified by sequential repetition. Nevertheless, this method and
its obvious variants break down when we consider the walk~$W(P)$ for
arbitrary chains~$P$, and arbitrary sets of marked elements.
Instead, we introduce a novel way to  approximate the diffusion operator.
Our approach is both conceptually simpler, and more general. We
observe that $W(P)$ amplifies the spectral gap of a
reversible Markov chain quadratically. We translate this to an
efficient approximation to the Grover diffusion operator
(\textbf{Theorem~\ref{thm-diffusion1}}), using the well known phase
estimation algorithm. 
We then begin an exposition of our algorithm by considering reversible Markov chains.  To
explain the basic idea of our approach, we first prove our main result
with an additional logarithmic factor
(\textbf{Theorem~\ref{thm-with-log-factor}}).

In Section~\ref{sec-search-approx}, using a technique developed by
H{\o}yer, Mosca, and de Wolf~\cite{HoyerMW03} we show how to eliminate
the logarithmic factor in the previous theorem, thus
proving \textbf{Theorem~\ref{thm-without-log-factor}}. 

In Section~\ref{sec-diffusion2}, we extend the algorithm to a
possibly non-reversible Markov chain whose discriminant
has non-zero singular value gap (\textbf{Theorem~\ref{thm-svg}}).
The complexity of the
algorithm in the general case is similar to the one for reversible
Markov chains.  The sole difference is that the singular value gap of
the discriminant matrix~$D(P)$
takes the place of the spectral gap of~$P$.  While the
eigenvalues of Markov chains are well studied, we are not aware of a
similar theory for singular values of this matrix. Nonetheless,
such a general result may prove useful for future applications.

\subsection{Subsequent work}

Since our work first appeared, much progress has been made on a question 
we had left unresolved. For any symmetric Markov chain,
Szegedy~\cite{Szegedy04} gave a procedure that detects the existence
of marked elements in time of the order of the square-root of the
classical hitting time. This result does not carry over to the
potentially harder problem of {\em finding\/} a marked element. The
latter (finding) problem has received particular attention in the case
of the $\sqrt{N} \times \sqrt{N}$ grid. The classical hitting time for
this graph is
in $\Order(N \log N)$
for any (non-zero) number of marked
elements.  Algorithms due to Ambainis, Kempe, and
Rivosh~\cite{AmbainisKR05} and Szegedy~\cite{Szegedy04} find a unique
marked state in time
$\Order(\sqrt{N} \, \log N)$,
a~$\sqrt{\log N}$ factor larger than the detection time.  In a recent paper,
Tulsi~\cite{Tulsi08} finally shows how we may find a unique marked
element in time $\Order(\sqrt{N \log N})$.

Magniez, Nayak, Richter, and Santha~\cite{MagniezNRS09} define
new, Monte Carlo type classical and quantum hitting times that are
potentially smaller than the existing notion of (Las Vegas type)
hitting times. They also present new quantum algorithms for the
detection and finding problems whose complexities are related to the
Monte Carlo quantum hitting time. The detection algorithm is based on
phase estimation, and the finding algorithm combines a similar phase
estimation based procedure with an idea introduced by Tulsi. Extending
Tulsi's result for the 2D grid, they show that for any
state-transitive Markov chain with a unique marked state, the quantum
hitting time is of the same order for both the detection and finding
problems.
Krovi, Magniez, Ozols, and Roland~\cite{kmor10} make a
significant improvement to this result by presenting a quantum algorithm 
for finding multiple marked elements in any reversible Markov chain. Taking a
new, simpler, and more general approach, they introduce a notion of
interpolation between any reversible chain and a perturbed version of
this chain, in which the marked states are absorbing.  The quantum
analogue of the interpolated walk not only detects but also finds
marked states with a quadratic speed-up over the classical hitting time.

\section{Quantum analogue of a classical Markov chain}
\label{sec-quantum-mc}
Let $P=(p_{xy})$ be the transition matrix of any irreducible Markov
chain on a finite space $X$ with $\size{X}=n$. 
We define a quantum analogue of 
$P$, based on and extending the notion of quantum Markov chain
due to Szegedy~\cite{Szegedy04}. The latter was inspired by an earlier
notion of quantum walk due to Ambainis~\cite{Ambainis04}. We also
point out that a similar process on regular graphs was studied by
Watrous~\cite{Watrous01}. 
Recall
that $P^*$ denotes the time-reversed Markov chain of $P$.

For a state~$\ket{\psi} \in \aitch$,
let~$\Pi_\psi = \ket{\psi}\bra{\psi}$ denote the orthogonal projector
onto~$\Span(\ket{\psi})$, and let $\reflex(\psi)= 2 \Pi_\psi - \id $
denote the reflection through the line generated by $\ket{\psi}$,
where~$\id$ is the identity operator on~${\aitch}$.
For a subspace~$\kay$
of $\aitch$ spanned by a set of mutually orthogonal
states $\{\ket{\psi_i} : i \in I\}$,
let $\Pi_{\kay} = \sum_{i \in I} \Pi_{\psi_i} $ be the orthogonal projector
onto $\kay$, and let 
$\reflex(\kay) = 2 \Pi_{\kay} - \id $
be the reflection through $\kay$.

Let $\ay=\Span( \ket{x}\ket{p_x} : x\in X)$ and $
\bee=\Span( \ket{p^*_y}\ket{y} : y\in X)$ be vector subspaces
of~$\aitch = \complex^{X \times X}$, where
\centermath{$
\ket{p_x} \quad =  \quad \sum_{y\in X} \sqrt{p_{xy}} \, \ket{y}
\quad\quad\textrm{and} \quad\quad
\ket{p^*_y}  \quad = \quad \sum_{x\in X} \sqrt{p^*_{yx}} \, \ket{x}.
$}
\begin{definition}[Quantum walk]
The unitary operation $W(P)$ defined on $\aitch$ by 
$W(P)  =  \reflex(\bee) \cdot \reflex(\ay)$
is called the {\em quantum walk\/} based on the classical chain~$P$.
\end{definition}
This quantum walk extends to a walk~$W(P)_d$ on the space~$\aitch$
augmented with data structures, as explained in
Section~\ref{sec-q-analogue}. Recall that~$\mathsf{U}$ is the quantum
update cost as defined in the same section.

\begin{proposition}\label{prop:wpd}
The quantum walk with
data, $W(P)_d$, can be implemented
at cost $4\mathsf{U}$.
\end{proposition}
\begin{proof}
Recall that $W(P)_d=\reflex(\bee)_d \cdot \reflex(\ay)_d$.
  The reflection~$\reflex(\ay)_d$ is implemented by mapping
  states~$\ket{x}_d\ket{p_x}_d$ to~$\ket{x}_d\ancilla_d$, applying
$\reflex (\ancilla_d)$ on the second register,
and
inverting
the first transformation. 
While the first and last steps each have cost $\mathsf{U}$, 
we only charge unit cost for the second step since it does not depend on the
data structure ($\ancilla_d=\ket{\bar{0},\bar{0}}$ by definition). Therefore the implementation of $\reflex(\ay)_d$ is of cost
$2\mathsf{U}$.  The reflection $\reflex(\bee)_d$ may be implemented
similarly.
\end{proof}

The eigen-spectrum of the transition matrix~$P$ plays an important
role in the analysis of a classical Markov chain. Similarly, the
behavior of the quantum process~$W(P)$ may be inferred from its
spectral decomposition. 
We consider the {\em
  discriminant} matrix~$D(P)=(\sqrt{p_{xy}p^*_{yx}})$. Since
$\sqrt{p_{xy}p^*_{yx}} = \sqrt{\pi_x} p_{xy}/ \sqrt{\pi_y}$, the
discriminant matrix is equal to
\centermath{
$D(P) \quad = 
\quad \diag(\pi)^{1/2} \cdot P \cdot \diag(\pi)^{-1/2},$
}
where~$\diag(\pi)$ is the invertible diagonal matrix with the
coordinates of the distribution~$\pi$ in its diagonal.  
Since the singular values of~$D(P)$ all lie in the range~$[0,1]$,
we may express them as~$\cos\theta$, for some angles~$\theta \in
[0,\tfrac{\pi}{2}]$.
(Note that this is a second type of use of the Greek letter~`$\pi$' in
this article, and it denotes the usual Mathematical constant. A third
type of use occurs later in the article. The meaning of the letter
can be inferred
from the context in which it is used.)
For later reference, we rewrite Theorem~1 
due to
Szegedy~\cite{Szegedy04}
which relates the singular value decomposition of~$D(P)$ to the 
spectral decomposition of~$W(P)$.
This theorem is a variant of a result due to Jordan~\cite{Jordan1875}
(see also Ref.~\cite[Section~VII.1, page~201]{Bhatia96}), and may be 
derived from it.
\begin{theorem}[Szegedy~\cite{Szegedy04}]
\label{thm-spectrum}
Let $P$ be an irreducible Markov chain, and
let $\cos\theta_1,\ldots,\cos\theta_l$ be an enumeration of those
singular values (possibly repeated) of $D(P)$ that lie in the open
interval~$(0,1)$.  Then:
\begin{enumerate}\reduceitemm

\item On $\ay + \bee$ those eigenvalues of $W(P)$
that have non-zero imaginary part are exactly $\e^{\pm 2 \complexi
\theta_1},\ldots,\e^{\pm 2 \complexi \theta_l}$, with the same
multiplicity.

\item On $\ay\cap\bee$ the operator $W(P)$ acts as the
identity~$\id$. 
The linear subspace~$\ay\cap\bee$
is spanned by
the left (and right) singular vectors of $D(P)$ with singular
value~$1$.

\item On $\ay \cap \bee^\perp$ and
$\ay^\perp\cap\bee$ the operator $W(P)$ acts as $-\id$.
The linear subspace~$\ay\cap\bee^\perp$ (respectively,
$\ay^\perp\cap\bee$) is spanned by
the set of left (respectively, right) singular vectors of $D(P)$ with
singular value~$0$.

\item $W(P)$ has no other eigenvalues on $\ay+\bee$;
on $\ay^\perp\cap\bee^\perp$ the operator $W(P)$ acts
as $\id$.
\end{enumerate}\reduceitemm
\end{theorem}

We define $\Delta(P)$, the {\em phase gap} of $W(P)$ as $2\theta$,
where $\theta$ is the smallest angle in $ (0,\tfrac{\pi}{2})$ such
that $\cos \theta$ is a singular value of $D(P)$.  This definition is
motivated by the previous theorem: 
in the complex plane,
the angular distance of $1$ from any other eigenvalue is at least
$\Delta(P)$.

\section{From quantum walk to search}
\label{sec-walk-search}

\subsection{Outline of search algorithm}
\label{sec-search}

We now describe a search algorithm that may be viewed as a quantum
analogue of \textbf{Search Algorithm~1} of
Section~\ref{sec-classical-search}.
Consider the following quantum state in the Hilbert space~$ \aitch$:
\[
\ket{\pi} 
    \quad = \quad \sum_{x \in X} \sqrt{\pi_x} \, \ket{x}\ket{p_x}
    \quad = \quad \sum_{y \in X} \sqrt{\pi_y} \, \ket{p_y^*}\ket{y}.
\]
(Note that its use as a label for the quantum state above is the third
type of use of the letter `$\pi$' in this article.)
This state serves as the initial state for our algorithm, and
corresponds to starting in the stationary distribution~$\pi$ in the
classical search algorithms.
Taking into account the data structure, preparing $\ket{\pi}_d$ from $\ancilla_d\ancilla_d$ has
cost $\mathsf{S} + \mathsf{U}$ as it requires one set-up operation to prepare $\sum_{x \in X} \sqrt{\pi_x} \, \ket{x}_d\ancilla_d$, followed by one update operation to map this state to $\ket{\pi}_d$.
Assume that 
$M \not= \emptyset$. 
Let $\emm=\complex^{M \times
X}$ denote the subspace with marked items in the first register.
We would like to transform the initial state~$\ket{\pi}$ to the target
state~$\ket{\mu}$, which is the (normalized) projection of~$\ket{\pi}$
onto the ``marked subspace'' $\emm$:
$$\ket{\mu} \quad = \quad \frac{\Pi_\emm\ket{\pi}}{\norm{\Pi_\emm\ket{\pi}}}
          \quad = \quad \frac{1}{\sqrt{\pM}} \sum_{x \in M} \sqrt{\pi_x} \,
                 \ket{x}\ket{p_x},
$$
where~$\pM=\norm{\Pi_\emm\ket{\pi}}^2 = \sum_{x \in M} \pi_x$
is the probability of a set~$M$ of marked states under the stationary
distribution~$\pi$. 
Roughly speaking, we effect this transformation by implementing a
rotation {\em \`a la\/} Grover~\cite{Grover96} in the two-dimensional
real subspace~$\ess = \Span(\ket{\pi}, \; \ket{\mu})$ generated by the
states.

Ideally, we would like to effect the rotation
$\reflex(\pi)_d \cdot \reflex(\mu^\perp)_d$
in
$\ess_d$,
where $\ket{\mu^\perp}$ is the state
in~$\ess$ orthogonal to~$\ket{\mu}$ which makes an acute angle
with~$\ket{\pi}$.
The angle~$\varphi$
between~$\ket{\pi}$ and~$\ket{\mu^\perp}$ is given by~$\sin \varphi =
\braket{\mu}{\pi} = \sqrt{\pM}$. The product of the two reflections
above is a rotation by an angle of~$2\varphi$ within the
space~$\ess$. Therefore, after~$\Order(1/\varphi) =
\Order(1/\sqrt{\pM})$ iterations of this rotation starting with the
state~$\ket{\pi}$, we
would
have approximated the target
state~$\ket{\mu}$.

Restricted to the subspace~$\ess$, the operators $\reflex(\mu^\perp)$ and
$- \reflex(\emm)$
are identical. Therefore, if we ensure that the state of the algorithm
remain
close to the subspace~$\ess$ throughout, we would be able to
implement $\reflex(\mu^\perp)_d$. This involves checking at cost 
$\mathsf{C}$ whether an item in the
first register is marked.

The reflection $\reflex(\pi)_d$ is
computationally harder to perform. The straightforward strategy would
be to rotate~$\ket{\pi}_d$ to the state~$\ket{\bar{0}}_d\ket{\bar{0}}_d$, use
$\reflex (\ket{\bar{0}}_d\ket{\bar{0}}_d)$, 
and then undo the first rotation. However, rotating~$\ket{\pi}_d$ to
~$\ket{\bar{0}}_d\ket{\bar{0}}_d$ is exactly the inverse operation of the preparation
of the initial state $\ket{\pi}_d$ from $\ket{\bar{0}}_d\ket{\bar{0}}_d$, and therefore
requires the same cost $\mathsf{S} + \mathsf{U}$. This may be much more expensive
than the update cost $4 \mathsf{U}$ incurred by the walk $W(P)_d$.
To use $W(P)_d$ instead,
our idea is to apply phase estimation to it, and exploit this procedure 
to approximate the required diffusion operator on $\ay_d + \bee_d$ which contains the
subspace~$\ess_d$.

\label{epsilon}
The above approach is only valid when the probability $\pM$ is known in
advance.  This assumption may be removed using standard techniques, 
without increasing the asymptotic complexity of the
algorithms~\cite{BoyerBHT98}. Indeed, if only a lower bound $\eps > 0$
on $\pM$ is known for non-empty~$M$, then the above argument 
can be modified in order to determine if $M$ is empty or find an 
element of $M$. We first sample from the stationary distribution a few
times to accommodate the case that~$\pM > 1/4$. If no marked element 
is found, we proceed as if~$\pM \leq 1/4$.
Following~\cite[Lemma~2]{BoyerBHT98},
we iterate the rotation $\reflex(\pi)_d \cdot \reflex(\mu^\perp)_d$
a total of~$T$ times on the initial state, where $T$ is chosen uniformly
at random in $[0,1/\sqrt{\eps}\,]$. If $M$ is not empty, a marked 
element is found with probability at least~$1/4$, and otherwise no marked 
element is found. We refer to this version of the Grover algorithm
as the {\em randomized Grover\/} algorithm.

\subsection{Diffusion operator from quantum walk}
\label{sec-diffusion1}

To explain our approach, in the rest of this section, and in the next
one, we assume that the classical Markov chain $P$ is ergodic and
reversible.  For a reversible chain the corresponding
discriminant~$D(P)$ is symmetric. Symmetry implies that the singular
values of~$D(P)$ equal the absolute values of its eigenvalues.
Since~$D(P) = \diag(\pi)^{1/2} \cdot P \cdot \diag(\pi)^{-1/2}$ is
similar to the matrix~$P$, their spectra are the same.  Therefore, we
only study the spectrum of~$P$.  The Perron-Frobenius theorem and
the ergodicity of~$P$ imply that the eigenvalue~$1$ has multiplicity~$1$,
and is the only eigenvalue of $P$ with absolute value $1$.
The corresponding eigenvector of~$D(P)$ is~$( \sqrt{\pi_x}\, )$, and
every singular (or eigen-) vector of~$D(P)$ orthogonal to this has
singular value strictly less than~$1$.  Transferring this property to
the quantum walk~$W(P)$ via Theorem~\ref{thm-spectrum}, 
$\ket{\pi}$ is the unique eigenvector of the unitary
operator~$W(P)$ in~$\ay + \bee$ with eigenvalue~$1$,
and the remaining eigenvalues in~$\ay + \bee$ are bounded away
from~$1$.  
We use this observation to identify
the component of any state~$\ket{\psi} \in \ess$ perpendicular
to~$\ket{\pi}$.

The main idea in our implementation of the above approach is to use 
phase estimation~$\cite{Kitaev95,KitaevSV02,CleveEMM98}$.
\begin{theorem}[Phase estimation; Cleve, Ekert, Macchiavello, and
Mosca~\cite{CleveEMM98}]
\label{thm-phase-estimation}
{}For every pair of integers $m,s\geq 1$, and 
a unitary operator~$U$ of
dimension~$2^m \times 2^m$,
there exists a quantum circuit $C(U)$ that acts on~$m+s$ qubits
and satisfies the following
properties:
\begin{enumerate}\reduceitemm

\item
The circuit~$C(U)$ uses $2s$~Hadamard gates, $\Order(s^2)$ controlled
phase rotations, and makes $2^{s+1}$ calls to the controlled unitary
operator $\control{U}$.

\item For any eigenvector~$\ket{\psi}$ of $U$ with eigenvalue~$1$,
i.e., if $U\ket{\psi}=\ket{\psi}$, then $C(U)\ket{\psi}\ket{0^s} =
\ket{\psi}\ket{0^s}$.
\item If $U\ket{\psi}=\e^{2\complexi \theta}\ket{\psi}$, where
$\theta\in(0,\pi)$, then $C(U)\ket{\psi}\ket{0^s} = \ket{\psi}\ket{\omega}$,
where $\ket{\omega}$ is an $s$-qubit state such that
$|\braket{0^s}{\omega}|=\sin(2^s\theta)/(2^s\sin\theta)$.
\end{enumerate}\reduceitemm
Moreover the family of circuits $C$ parametrized by~$m$ and $s$ is
uniform. 
\end{theorem}
This circuit is called phase estimation because measuring the 
state
$\ket{\omega}$ in the computational basis 
yields an approximation to~$\theta/\pi$.
In our case we only need to discriminate between the 
eigenvalue $1$ and the remaining eigenvalues.
In the following theorem we show how phase estimation
is
used to design a quantum circuit $R(P)$ which implements an operation that
is close to the reflection $\reflex(\pi)$.

\begin{theorem}
\label{thm-diffusion1}
Let~$P$ be an ergodic Markov chain on a state space of
size~$n \geq 2$, such that the phase gap of the quantum walk~$W(P)$
based on~$P$ is~$\Delta(P)$.
Then
for any integer~$k$
there exists a quantum circuit $R(P)$ that
acts on~$2 \lceil{\log_2 n \rceil} + ks$  qubits,
where~$s \in \log_2(\frac{1}{\Delta(P)})+\Order(1)$,
and satisfies the following properties:
\begin{enumerate}\reduceitemm

\item
The circuit~$R(P)$ uses $2ks$~Hadamard gates, $\Order(k s^2)$
controlled phase rotations, and makes at most $k\, 2^{s+1}$ calls to
the controlled quantum walk $\control{W(P)}$ and its
inverse~$\control{W(P)}^\adjoint$.

\item If~$\ket{\pi}$ is the unique~$1$-eigenvector of~$W(P)$ as
defined above, then~$R(P)\ket{\pi}\ket{0^{ks}} = \ket{\pi}\ket{0^{ks}}$.

\item If~$\ket{\psi}$ lies in the subspace of~$\ay + \bee$
orthogonal to~$\ket{\pi}$,
then~$\norm{ (R(P) + \id) \ket{\psi}\ket{0^{ks}} } \leq 2^{1-k}$.

\end{enumerate}\reduceitemm
Moreover the family of circuits $R(P)$ parametrized by~$n$ and $k$ is
uniform. 
\end{theorem}
\begin{proof}
We describe the circuit $R(P)$.
Let $m=n^2$ and $s=\left\lceil\log_2(\tfrac{2\pi}{\Delta(P)})\right\rceil$.
We start by applying the phase estimation circuit $C(U)$ to the quantum walk
$W(P)$, a unitary operator of dimension~$m \times m$. To increase the 
accuracy of the phase estimation, we repeat the circuit $k$~times, 
creating $k$ identical copies of the $s$-qubit state~$\ket{\omega}$ 
holding estimates of the phase.
Observe that only the number of ancillary qubits increases from~$s$
to~$ks$ in this process. Since $C(U)$ leaves the eigenvectors of $W(P)$ 
in the first register unchanged, we do not need additional copies of the 
state~$\ket{\psi}$.

The above operations approximately resolve any state
$\ket{\psi}$ in $\ay + \bee$
along the eigenvectors of~$W(P)$ by labeling them with estimates of
the corresponding eigenvalue phases.  We now flip the phase (i.e.,
multiply it by~$-1$) of all
computational basis states with a non-zero estimate of the phase in any
of the~$k$ copies. Our intention is to flip the phase
of all eigenvectors other than~$\ket{\pi}$. Finally, we reverse the
phase estimation. All these operations together constitute~$R(P)$.

The state $\ket{\pi}\ket{0^{ks}}$ stays unchanged
under the action of~$R(P)$.  
When $\ket{\psi}$ is orthogonal to $\ket{\pi}$
it is a linear
combination of eigenvectors of $W(P)$ whose eigenvalues are of the
form $\e^{\pm 2 \complexi \theta}$, where $\Delta(P)/2 \leq \theta <
\pi/2$.
By definition of $s$, the 
state $\ket{\omega}$ holding the estimate 
for any phase $\theta\neq 0$ then satisfies
$|\braket{0^s}{\omega}|\leq 1/2$.
With
$k$ repetitions of the phase estimation, we can therefore decompose
$\ket{\psi}\ket{0^{ks}}$ into a sum $\ket{\psi_0} + \ket{\psi_1}$, 
such that the phase estimate is zero in each of the~$k$ copies
of~$\ket{\omega}$ on the state $\ket{\psi_0}$, is non-zero in at least 
one copy on the state $\ket{\psi_1}$, and
$\norm{\psi_0} \leq 2^{-k}$.
Then $R(P) \ket{\psi}\ket{0^{ks}} = \ket{\psi_0} -
\ket{\psi_1}$, and $(R(P) + \id) \ket{\psi}\ket{0^{ks}} =
2\ket{\psi_0}$, whose norm is at most
$2^{1-k}$.
\end{proof}

\subsection{The search algorithm for reversible Markov chains}

Let us consider the following quantum procedure.
\encadre{ $\textbf{Quantum Search}(P,\eps)$
\begin{enumerate}\reduceitem
\item\label{step-pM}
Repeat 5 times:
\begin{enumerate}\reduceitem
\item Sample a state~$x$ from the stationary distribution~$\pi$ of~$P$.
\item If $x\in M$, output $x$ and stop.
\end{enumerate}

\item Choose $T$ uniformly at random in $[0,1/\sqrt{\eps}\,]$, let~$k
  \in \log_2(T)+\Order(1)$, and let~$s$ be as given by
  Theorem~\ref{thm-diffusion1}.

\item 
Prepare the initial state $\ket{\pi}_d \ket{0^{Tks}}$.

\item\label{step:grover-iterations} Repeat $T$ times:
\begin{enumerate}\reduceitemm
\item For any basis vector $\ket{x}_d\ket{y}_d \ket{z}$ of~$\aitch_d$
  and the ancillary~$(Tks)$-qubit space, flip the phase if $x\in M$:
\[
\ket{x}_d\ket{y}_d \ket{z} \mapsto
    \left\{ \begin{array}{rl}
        - \ket{x}_d\ket{y}_d \ket{z}, & \quad \textrm{if } x \in M \\
        \ket{x}_d\ket{y}_d \ket{z},   & \quad \textrm{otherwise.}
        \end{array}
    \right.
\]

\item Apply circuit $R(P)_d$ of Theorem~\ref{thm-diffusion1}
  with $k$ as above, using a fresh set of ancilla
  qubits~$\ket{0^{ks}}$ in each iteration.
\end{enumerate}
\item Observe the first register.
\item Output $x$ if $x\in M$, otherwise output `no marked element
  exists'.
\end{enumerate}
} 
\begin{theorem}
\label{thm-with-log-factor}
Let~$\delta > 0$ be the eigenvalue gap of a reversible, ergodic Markov
chain~$P$, and let $\eps > 0$ be a lower bound on the probability that
an element chosen from the stationary distribution of $P$ is marked
whenever~$M$ is non-empty.  Then, with high probability, 
the procedure \textbf{Quantum Search$(P,\eps)$}
determines if $M$ is empty or
else
finds an element of $M$ with cost of order
$
\mathsf{S} + \frac{1}{\sqrt{\eps}} \left[
\left( \frac{1}{\sqrt{\delta}}\log\frac{1}{\sqrt{\eps}} \right)
\mathsf{U}+\mathsf{C} \right].
$
\end{theorem}

\begin{proof}
{}For convenience,
we reason in the Hilbert space $\mathcal{H}$,
without the data structures, and also
omit the ancilla qubits used by the circuit~$R(P)$. Between 
applications of $R(P)$ the ancilla qubits remain in a state close 
to $\ket{0^{Tks}}$.

First observe that if $M$ is empty then no marked element is found by 
\textbf{Quantum Search$(P,\eps)$}.  We assume now that $M$ is non-empty.
When~$\pM > 1/4$, we detect a marked element in Step~\ref{step-pM} with
probability at least~$1 - (3/4)^5 > 3/4$. In analyzing the correctness
of the remaining steps, we may therefore assume that~$\pM \leq 1/4$.
Let~$\ess$ be the two-dimensional subspace~$\ess = \Span(\ket{\pi}, 
\; \ket{\mu})$. Recall that the randomized Grover algorithm consists
in $T$ iterations of $\reflex(\pi)\cdot\reflex(\mu^\perp)$,
where~$T$ is chosen uniformly at random from $[0,1/\sqrt{\eps}\,]$.
Since $\eps\leq \pM \leq 1/4$, with constant probability the randomized
Grover algorithm rotates the vector $\ket{\pi}$ in the space~$\ess$
into a state whose inner product with $\ket{\mu}$ is a constant.
Using a hybrid argument as in Refs.~\cite{BennettBBV97,Vazirani98b}, we 
prove that the algorithm \textbf{Quantum Search$(P,\eps)$} simulates, 
with an arbitrarily small
constant probability of 
error, the
randomized
Grover algorithm, and therefore finds a marked element with high probability,
whenever such an element exists. 

{}For $i\geq0$, we define $\ket{\phi_i}$ as the result of
$i$ Grover iterations applied to $ \ket{\pi}$, and 
$\ket{\psi_i}$ as the result of
$i$  iterations of step (\ref{step:grover-iterations}) in 
\textbf{Quantum Search$(P,\eps)$} applied to $ \ket{\pi}$. We show by induction on $i$, that
$\norm{\ket{\psi_i} - \ket{\phi_i}} \leq i 2^{1-k}$.
Indeed, we can write $\ket{\psi_i}$ as
$ \ket{\phi_i} + (\ket{\psi_i} - \ket{\phi_i})$. 
The actions of  $\reflex(\mu^\perp)$ and  $- \reflex(\emm)$ are
identical on $ \ket{\phi_i}$  since the state is in $\ess$. Set 
$\ket{\tau} = 
\ket{\phi_{i+1}} - R(P)\cdot \reflex(\emm) \ket{\phi_i}$.
Since $\reflex(\emm) \ket{\phi_i}$ is in $\ess$, and $\ess$ is a
subspace of $\ay + \bee$, conclusion (3) of
Theorem~\ref{thm-diffusion1} can be applied, which implies that
$\norm{\tau} \leq 2^{1-k}$.
Using $\norm{\ket{\psi_{i+1}} -
  \ket{\phi_{i+1}}} \leq \norm{\ket{\tau}} + \norm{\ket{\psi_i} -
  \ket{\phi_i}}$, the statement follows.
For $k \in \log_2(T)+c$, where $c$ is a constant, this implies that
$\norm{\ket{\psi_T} - \ket{\phi_T}}\leq 2^{1-c}$, which can be made arbitrarily small by choosing $c$ sufficiently large.

Let us now turn to the cost of the procedure.
Since measuring~$\ket{\pi}$
gives us a sample from the stationary distribution~$\pi$, the cost of
Step~\ref{step-pM} is of the order of~$\mathsf{S}$.
Preparing~$\ket{\pi}_d$ costs
$\mathsf{S} + \mathsf{U}$, and in each iteration the single phase flip
costs $\mathsf{C}$. In the circuit $R(P)_d$, the controlled quantum
walk and its inverse can be implemented with four update operations,
each of cost $\mathsf{U}$.  Indeed, the implementation of $W(P)$,
described in the proof of Proposition~\ref{prop:wpd} works also for
the controlled quantum walk if we replace
$\reflex ( \ancilla_d )$ 
by the controlled operator
$\control{\reflex ( \ancilla_d )}$.
Since the controlled reflection is also of unit cost, this change does
not alter the cost of the implementation.

In $R(P)_d$ the number of controlled quantum walks and its inverse is
in $\Order( (1/{\Delta(P)}) \log ( 1/ \sqrt{\epsilon}) )$. We claim
that $\Delta(P) = \Omega ( \sqrt{\delta})$.
Let~$\lambda_0,...,\lambda_{n-1}$ be the
eigenvalues of $P$, possibly with repetitions,
such that $1 =\lambda_0 > \abs{\lambda_1} \geq \ldots \geq
\abs{\lambda_{n-1}}$.  Since the discriminant~$D(P)$ is similar
to~$P$, their spectra are the same, and therefore the singular values
of $D(P)$ are~$\abs{\lambda_0}, \abs{\lambda_1}, \dotsc,
\abs{\lambda_{n-1}}$.  By definition, $\Delta(P) =2 \theta_1$,
where~$\cos \theta_1 = \abs{\lambda_1}$.
The following straightforward (in)equalities 
relate~$\Delta(P)$ to~$\delta(P)$:
$ \Delta(P)
     \geq  \abs{1 - \e^{2 \complexi \theta_1 }}
     =  2 \sqrt{1 - \abs{\lambda_1}^2}
    \geq  2\sqrt{\delta}.
$
This finishes the cost analysis.
\end{proof}

Let us observe that the origin of the quadratic speed-up due to
quantum walks may be traced to the quadratic relationship between the
phase gap~$\Delta(P)$ of the quantum walk~$W(P)$ and the eigenvalue
gap~$\delta$ of the classical Markov chain~$P$, observed at the end of
the
above
proof.

\section{Search with approximate reflection operators}
\label{sec-search-approx}

In this section, we describe how our approximate reflection operator
may be incorporated into a search
algorithm without incurring additional cost for reducing its
error. 
The basic idea is to adapt the recursive amplitude amplification ({RAA}) algorithm
due to H{\o}yer, Mosca, and de~Wolf~\cite{HoyerMW03} to our
setting. To describe it, we use the notation from
Section~\ref{sec-search} where we discussed how the Grover algorithm
works to rotate a starting state $\ket{\pi}$ into a target state
$\ket{\mu}$, where $\braket{\mu}{\pi} = \sin \varphi =
\sqrt{\pM}$. We define procedures $A_i$ recursively, for~$i
\geq 0$.  Let the procedure~$A_0$ be the identity map~$\id$, and for
$i > 0$, let
$$
A_i \quad = \quad  A_{i-1} \cdot \reflex(\pi) \cdot A_{i-1}^\adjoint \cdot 
\reflex(\mu^\perp) \cdot A_{i-1}.
$$
We define the states $\ket{\pi_i}$ as $A_i \ket{\pi}$. Then
$\ket{\pi_i}$ forms an angle $3^i \varphi$ with $\ket{\mu^\perp}$, and
therefore the state~$\ket{\pi_t}$ is close to~$\ket{\mu}$ when~$t =
\log_3\frac{1}{\varphi} + \Order(1)$. The final recursive algorithm is
thus~$A_t$.

We may estimate the cost~$\mathsf{Cost}(t)$ of this search algorithm
in terms of the cost~$c$ of implementing the two original
reflections, $\reflex(\pi)$ and~$\reflex(\mu^\perp)$. We
have~$\mathsf{Cost}(0) = 0$, and for~$i \geq 1$, $ \mathsf{Cost}(i) =
3 \cdot \mathsf{Cost}(i-1) + c,$ and therefore the cost of $A_t$
is $\Order ( c / \sqrt{\varepsilon})$.

The RAA algorithm is more suitable for
situations where we have imperfect procedures that implement the basic
reflections~$\reflex(\pi), \reflex(\mu^\perp)$. H{\o}yer {\it et
al.\/}~\cite{HoyerMW03} demonstrated this when there is an ideal
(error-free) procedure for~$\reflex(\pi)$, and a procedure
for~$\reflex(\mu^\perp)$ that has ideal behavior only with high
probability.
Here, we adapt their approach to the case where it is the first reflection $\reflex(\pi)$ which may only be approximated (it is probably possible to deal with the case where both reflections are imperfect, but for the sake of simplicity, we only deal with the case when the implementation of $\reflex(\mu^\perp)$ is ideal since this is sufficient for our purpose).
In the context of quantum walk based
search, an imperfection appears in the form given by
Theorem~\ref{thm-diffusion1}. The basic idea is to 
create an analogue of the recursive algorithms $A_i$ when 
$\reflex(\pi)$ is replaced by increasingly fine approximations based on
Theorem~\ref{thm-diffusion1}.

We now state this precisely in full generality for potential further applications.
Assume that for any $\beta > 0$, we have a quantum circuit
$R(\beta)$ acting on $\aitch\otimes\kay$, where $\kay$ is an extra
register of $s(\beta)$ qubits.
For a given integer $t$, and a precision parameter $\gamma$,
the quantum circuit consists of $t$ induction steps 
and acts on $\aitch \tensor \left[ \bigotimes_{i=1}^t \kay_i
\right]$, where $\kay_i$ is an extra register used at step $i$. Let
$s_i = s(\beta_i)$ be the size of register $\kay_i$,
Let~$S=\sum_{i=1}^t s_i$.
We use $\ket{\pi}_d\ket{0^S}$ as the initial state of the algorithm.

The quantum circuit follows exactly the RAA algorithm explained above. We essentially replace 
$\reflex(\pi)$ at step $i$ by an approximation $R(\beta_i)$, acting on~$\aitch \tensor \kay_i$, and
$\id$ on the rest.  Here is now one explicit step of the induction, where
the basis case $\textbf{Approximate RAA}(0,\gamma)$ is simply the identity map~:
\encadre{ $\textbf{Approximate RAA}(i,\gamma)$
\begin{enumerate}\reduceitem
\item Apply $\textbf{Approximate RAA}(i-1,\gamma)$.
\item For any basis vector $(\ket{x}_d\ket{y}_d)\otimes \ket{z}$, where $\ket{x}\ket{y}\in\aitch$, flip the phase  if $x\in M$.
\item Undo $\textbf{Approximate RAA}(i-1,\gamma)$.\label{rec-start}
\item \label{conditioning} If any of the registers~$\kay_j$, with~$j < i$, are not in state~$\ket{0^{s_j}}$, respectively, then flip the
phase of the state. Otherwise, apply $R(\beta_i)$ on $\aitch\otimes\kay_i$,
where  $\beta_i=\tfrac{18}{4\pi^3}\gamma/i^2$.
\item Apply $\textbf{Approximate RAA}(i-1,\gamma)$.\label{rec-end}
\end{enumerate}
} 

We now prove that this algorithm can be used to find a marked element when $\pM$ is known. We will later show how to modify the algorithm when only a lower bound on $\pM$ is known.
\begin{lemma}\label{lemma-grover-approximate-reflection}
Assume that for any $\beta > 0$, we have a quantum circuit
$R(\beta)$ acting on $\aitch\otimes\kay$, where $\kay$ is an extra
register of $s$ qubits ($s = s(\beta)$ may depend on $\beta$), with
the following properties:
\begin{enumerate}\reduceitemm
\item
The circuit~$R(\beta)$ has cost $c_1 \log \tfrac{1}{\beta}$.

\item $R(\beta)\ket{\pi}\ket{0^{s}} = \ket{\pi}\ket{0^{s}}$.

\item 
$\norm{ (R(\beta) + \id) \ket{\psi}\ket{0^{s}} } \leq \beta$
when $\ket{\psi}$ is orthogonal to~$\ket{\pi}$.

\end{enumerate}\reduceitemm
Further, assume that we are able to apply 
$-\reflex ( \emm )$ with cost $c_2$,
and let~$t$ be the smallest non-negative integer such that $3^t \,\sin^{-1}\sqrt{\pM} \quad \in \quad \left[{\pi}/{4}, {3\pi}/{4}
\right]$.  
Then, for every real~$\gamma > 0$, 
$\textbf{Approximate RAA}(t,\gamma)$ maps $\ket{\pi}\ket{0^S}$ 
to a state that has projection of length at least
$(\tfrac{1}{\sqrt{2}}- \gamma)$ in $\emm\otimes\left[ \bigotimes_{i=1}^t \kay_i \right]$,
and incurs a cost of order
$3^t \cdot (c_1 \log\tfrac{1}{\gamma} + c_2)$.
\end{lemma}
\begin{proof}
For simplicity, we omit the data structure in our error analysis,
but take it into account in bounding the complexity of the algorithm.
Let $s_i = s(\beta_i)$ be the size of register $\kay_i$.
Let~$S=\sum_{i=1}^t s_i$.
Recall
that  we use $\ket{\phi_0} = \ket{\pi}\ket{0^S}$
as the initial state. We also denote by $\ket{\phi_i}$ the output state   of 
$\textbf{Approximate RAA}(i,\gamma)$ on input
$\ket{\phi_0}$.
Note that the component of $\ket{\phi_i}$ on $\kay_j$ is $\ket{0^{s_j}}$, for all $j>i$.
Define the reflection operator~$R_i$ as the product of the recursive steps~\ref{rec-start}-\ref{rec-end}
of $\textbf{Approximate RAA}(i,\gamma)$.

In order to understand the behavior of $R_i$,
let us examine the action of $R(\beta_i)$ in step~\ref{conditioning}.
At the beginning of that step, the algorithm state still has component $\ket{0^{s_j}}$ on $\kay_j$, for all $j\geq i$. 
Therefore the conditioning, and the fact that
$R(\beta_i)$ is an approximation to $\reflex(\pi)$, directly gives
that $R_{i}$ behaves on the current state as an approximation to $\reflex(\phi_{i-1})$. 
To be more precise, let $E_i =R_i-\reflex(\phi_{i-1})$ be the error made in our implementation
of~$\reflex(\phi_{i-1})$. We state the following fact without proof 
since it directly derives from the hypothesis on $R(\beta_i)$.
\begin{fact}
\label{fact-error-properties}
$E_i$ satisfies the following properties:
\begin{enumerate}\reduceitemm

\item \label{fact-prop1}
$E_i \ket{\phi_{i-1}} = 0$, and

\item
\label{fact-prop2}
$\norm{E_i\ket{\psi}\ket{0^{S_i}}} \leq \beta_i$,
for all~$\ket{\psi}\in\aitch \tensor \left[ \bigotimes_{j = 1}^{i-1} \kay_j \right]$
such that $\ket{\psi}\ket{0^{S_i}} \perp \ket{\phi_{i-1}}$,
where $S_i=\sum_{j=i}^t s_j$.
\end{enumerate}\reduceitemm
\end{fact}

To analyze this algorithm, we keep track 
of the projection of $\ket{\phi_i}$ on the marked subspace. The marked
subspace corresponds to~$\emm \tensor \left[ \bigotimes_j \kay_j
\right]$; it consists of states in which the first register of the~$\aitch$-part is
marked.
We denote this space by~$\emmt$.  Define
the normalized projections of $\ket{\phi_i}$ on the marked subspace
$\emmt$ and on its orthogonal complement as:
\begin{eqnarray*}
\ket{\mu_i} 
    & = & \frac{\Pi_\emmt \ket{\phi_i}}{\norm{\Pi_\emmt \ket{\phi_i}}} \\
\ket{\mu_i^\perp} 
    & = & \frac{(\id-\Pi_\emmt) \ket{\phi_i}}
               {\norm{(\id-\Pi_\emmt) \ket{\phi_i}}}.
\end{eqnarray*}
We thus have
\begin{equation}
\label{eq:phik}
\ket{\phi_i} \quad = \quad 
    \sin\varphi_i \; \ket{\mu_i} + \cos\varphi_i \; \ket{\mu_i^\perp}.
\end{equation}
where $\sin^2\varphi_i=\norm{\Pi_\emmt\ket{\phi_i}}^2$ is
the probability of finding a marked
item by measuring the first register according to~$\set{\Pi_\emmt, \id - \Pi_\emmt}$.  
For later use, let us also define $\ket{\phi_i^\perp}$ as the state in the $2$-dimensional
subspace spanned by $\ket{\mu_i}$ and $\ket{\mu_i^\perp}$ that is
orthogonal to $\ket{\phi_i}$:
\begin{equation*}
\ket{\phi_i^\perp} \quad = \quad 
    \cos\varphi_i \; \ket{\mu_i} - \sin\varphi_i \; \ket{\mu_i^\perp}.
\end{equation*}

For the initial state $\ket{\phi_0}$, we have $\sin^2\varphi_0=p_M$. 
If all the errors $\beta_i$ were
zero, $\textbf{Approximate RAA}$ would implement the RAA algorithm 
in the subspace spanned by $\ket{\mu_i} = \ket{\mu_0}$
and $\ket{\mu_i^\perp} = \ket{\mu_0^\perp}$, with the angles~$\varphi_{i+1}=3{\varphi}_i$, 
that is~$\varphi_i =3^i\varphi_0$. 
Therefore by recursively iterating our procedure for a total number
of $t$ steps, we would end up with a state whose inner product with $\ket{\mu_0}$ is
at least~$\tfrac{1}{\sqrt{2}}$.

\textbf{Analysis of the errors ---}
We show that $\textbf{Approximate RAA}$ still works when the errors $\beta_i$ are sufficiently small.  
In that case, the $2$-dimensional subspace $\Span(\ket{\mu_i}, \ket{\mu_i^\perp})$ may drift away from
the initial subspace $\Span( \ket{\mu_0}, \ket{\mu_0^\perp})$, and the
angles~$\varphi_i$ may be different from the ideal
value~$\bar{\varphi}_i = 3^i\varphi_0$. We derive bounds on the
error~$e_i$:
\begin{equation}
\label{eqn-error}
e_i \quad = \quad \size{\sin\varphi_i-\sin\bar{\varphi}_i},
\end{equation}
the difference between the amplitude~$\sin \varphi_{i}$ of the marked
part of the state~$\ket{\phi_{i}}$ and the ideal amplitude,~$\sin
\bar{\varphi_i}$.

We assume without loss of generality that 
$0 < \gamma < \tfrac{1}{\sqrt{2}}$ since the case
$\gamma \geq \tfrac{1}{\sqrt{2}}$ is vacuous. We prove that after $t$ steps $e_t \leq \gamma$.
This will conclude the error analysis 
since $\frac{1}{\sqrt{2}}\leq\sin \bar{\varphi}_t \leq 1$.

We have
\begin{eqnarray}
\ket{\phi_{i+1}} 
    & = & R_{i+1} \cdot \reflex(\emmt^\perp) \; \ket{\phi_i} \nonumber \\
    & = & \reflex(\phi_i) \cdot \reflex(\emmt^\perp) \; \ket{\phi_i}
          + E_{i+1} \cdot \reflex(\emmt^\perp) \; \ket{\phi_i} \nonumber \\
\label{eq:phikplus1}
    & = & \sin 3\varphi_i \; \ket{\mu_i} 
          + \cos 3\varphi_i \; \ket{\mu_i^\perp}
          + \ket{\omega_{i+1}},
\end{eqnarray}
where we used the fact that $\reflex(\phi_i) \cdot \reflex(\emmt^\perp)$ implements a
perfect amplitude amplification step, and we introduced an error
state $\ket{\omega_i}$, defined as
\begin{eqnarray*}
\ket{\omega_{i+1}}
    & = & E_{i+1} \cdot \reflex(\emmt^\perp) \; \ket{\phi_i}\\
    & = & E_{i+1} \cdot \reflex(\emmt^\perp)
         \left(\sin\varphi_i \, \ket{\mu_i} 
             + \cos\varphi_i \, \ket{\mu_i^\perp} \right) \\
    & = & E_{i+1} \left( -\sin\varphi_i \, \ket{\mu_i} 
          + \cos\varphi_i \, \ket{\mu_i^\perp} \right) \\
    & = & E_{i+1} \left( \cos 2\varphi_i \, \ket{\phi_i} 
          - \sin 2\varphi_i \, \ket{\phi_i^\perp} \right) \\
    & = & -\sin 2\varphi_i \;  E_{i+1} \, \ket{\phi_i^\perp},
\end{eqnarray*}
where we used Fact~\ref{fact-error-properties},
property~\ref{fact-prop1}. Moreover,
$\ket{\phi_i^\perp} \perp \ket{\phi_i}$, so
$\norm{\omega_{i+1}} \leq \beta_{i+1} \size{ \sin 2\varphi_i }$ by Fact~\ref{fact-error-properties}, property~\ref{fact-prop2}.
Finally, comparing Eq.~(\ref{eq:phik}) and Eq.~(\ref{eq:phikplus1}), we
get
\begin{equation*}
|\sin\varphi_{i+1}-\sin 3\varphi_{i}| 
    \quad \leq \quad \beta_{i+1} \size{\sin 2\varphi_i}.
\end{equation*}

We may now bound the error defined in Eq.~(\ref{eqn-error}) as:
\begin{eqnarray}
\nonumber
e_{i+1} & \leq & |\sin\varphi_{i+1}-\sin 3\varphi_{i}| 
                 + |\sin3\varphi_i-\sin\bar{\varphi}_{i+1}| \\
\nonumber
        & \leq & \beta_{i+1} \size{ \sin 2\varphi_i }
                 + |\sin3\varphi_i - \sin3\bar{\varphi}_{i}| \\
\nonumber
        & \leq & \beta_{i+1} (\sin 2\bar{\varphi}_i + 
                 |\sin2\varphi_i - \sin2\bar{\varphi}_{i}|) \\
\nonumber
        &      & ~~~~ + ~ |\sin3\varphi_{i} - \sin3\bar{\varphi}_{i}| \\
\nonumber
        & \leq & \beta_{i+1} (\sin 2\bar{\varphi}_i+ 2e_i) + 3 e_i \\
\label{eqn-omega}
        & \leq & 2\beta_{i+1} (\bar{\varphi}_i+ e_i) + 3 e_i,
\end{eqnarray}
where we have used the triangle inequality and the following trigonometric
inequalities
\begin{eqnarray*}
|\sin 2A - \sin 2B| & \leq & 2|\sin A-\sin B|\\
|\sin 3A - \sin 3B| & \leq & 3|\sin A-\sin B|\\
\sin A & \leq & A
\end{eqnarray*}
that hold for any angles $A,B\in[0,\pi/4]$.

We define a quantity~$\tilde{e}_i$, intended to be an
upper bound on $e_i$ (it would be if $\tilde{e}_i\leq\bar{\varphi}_i$). Let
\begin{eqnarray*}
\tilde{e}_0&=&0\\
\tilde{e}_{i+1}&=& 4\beta_{i+1} \bar{\varphi}_i + 3 \tilde{e}_i.
\end{eqnarray*}
We show that $\tilde{e}_i \leq \gamma$ for every~$i \le t$.
Indeed, let us define $u_i$ as
\begin{equation*}
\tilde{e}_i \quad = \quad \gamma\, \bar{\varphi}_i\, u_i.
\end{equation*}
We therefore have the following recursion for $u_i$
\begin{eqnarray*}
u_0&=&0\\
u_{i+1}&=&u_i+\frac{4}{3\gamma}\beta_{i+1}, \quad\quad (\forall i \geq 0)
\end{eqnarray*}
so that
\begin{equation*}
u_i \quad = \quad \frac{4}{3\gamma}\sum_{j=1}^i\beta_j.
\end{equation*}
Recall that we have chosen $\beta_i=\tfrac{18}{4\pi^3}\gamma/i^2$, so that 
$\set{\beta_i}$ define a convergent series and
the non-decreasing sequence $(u_i)$ tends to $1/ \pi$ when $i\to\infty$.
We therefore have
%
$\tilde{e}_i \leq \gamma \bar{\varphi_i}/\pi
\leq \gamma$ 
since
$ 0 \leq \  \bar{\varphi_t} \leq \pi$
for $i \leq t$.

Since $0<\gamma\leq 1$, we have 
$\tilde{e}_i \leq \bar{\varphi_i}$, and
we can show by induction that~$e_i \leq \tilde{e}_i$
for all~$i \leq t$. This finishes the error analysis.

\textbf{Complexity ---} We now evaluate the complexity of our
algorithm. We know from the hypotheses of the theorem that applying
$R(\beta_i)$ costs $c_1 \log\tfrac{1}{\beta_i}$, while applying
$\reflex(\emmt^\perp)= - \reflex(\emm) \tensor \id_{\bigotimes_j \kay_j}$ costs $c_2$.
Moreover, by definition of $\textbf{Approximate RAA}$,
applying $\textbf{Approximate RAA}(i,\gamma)$ requires $3$ calls to
$\textbf{Approximate RAA}(i-1,\gamma)$, 
one call to $R(\beta_i)$ and one call to $\reflex(\emm)$.
Hence, if we denote by $\mathsf{Cost}(i)$ the cost of applying $\textbf{Approximate RAA}(i,\gamma)$,
we have
\begin{eqnarray*}
\mathsf{Cost}(0)&=&0\\
\mathsf{Cost}(i)&=&3 \, \mathsf{Cost}(i-1) + c_1 \log\frac{1}{\beta_i} + c_2.
\end{eqnarray*}
Since we have fixed $\beta_i=\tfrac{18}{4\pi^2}\gamma/i^2$,
we find that~$\mathsf{Cost}(i)$ equals
\begin{eqnarray*}
\lefteqn{
    c_1 \sum_{j=1}^i 3^{i-j} \left(2 \log j + \log \frac{1}{\gamma}+\Order(1)\right)          
          + c_2 \sum_{j=1}^i 3^{i-j}
} \\ 
    & = & 3^i \left[ \left( c_1\log \frac{1}{\gamma}+c_2+\Order(1)\right)
          \sum_{j=1}^i \frac{1}{3^j} 
          + 2c_1 \sum_{j=1}^i \frac{\log j}{3^j} \right]
\end{eqnarray*}
where both sums converge as $i\to\infty$. After
$t$
steps we have $\mathsf{Cost}(t) \in \Order\left(
3^t \cdot ( c_1 \log\frac{1}{\gamma} + c_2) \right)$.

If the cost refers to time complexity, then there is an additional
term pertaining to the reflection~$R_i$. This arises from the check to
see if the ancilla are in state~$\ket{0^S}$. This does not change
the asymptotic complexity of the algorithm.
\end{proof}

Note that Lemma~\ref{lemma-grover-approximate-reflection}
requires knowledge of $\pM$
to infer the necessary number of iterations $t$. When only a lower bound $\eps$ on $\pM$ is known, we can use the algorithm $\textbf{Tolerant RAA}(t,\gamma)$, which only adds a 
constant factor overhead
with respect to $\textbf{Approximate RAA}(t,\gamma)$.

\encadre{ $\textbf{Tolerant RAA}(\tmax,\gamma)$
\begin{enumerate}\reduceitem
\item Sample a state $x$ from the stationary distribution $\pi$ of $P$.
\item if $x\in M$, output $x$, and stop.
\item Prepare the initial state $\ket{\pi}_d\ket{0^S}$ and set $i=0$.
\item\label{step:new-attempt} Increment $i$. Apply $\textbf{Approximate RAA}(i,\gamma)$.
\item Measure the first register according to $\Pi_\emm$.\\
If successful, observe and output the first register, and stop. 
\item If $i< \tmax$ go back to Step~\ref{step:new-attempt}, otherwise output ``No marked element''.
\end{enumerate}
} 

\begin{lemma}\label{lemma-unknown-pm}
Assume that for any $\beta > 0$, we have a quantum circuit
$R(\beta)$ acting on $\aitch\otimes\kay$, where $\kay$ is an extra
register of $s$ qubits ($s = s(\beta)$ may depend on $\beta$), with
the following properties:
\begin{enumerate}\reduceitemm
\item
The circuit~$R(\beta)$ has cost $c_1 \log \tfrac{1}{\beta}$.

\item $R(\beta)\ket{\pi}\ket{0^{s}} = \ket{\pi}\ket{0^{s}}$.

\item 
$\norm{ (R(\beta) + \id) \ket{\psi}\ket{0^{s}} } \leq \beta$
when $\ket{\psi}$ is orthogonal to~$\ket{\pi}$.

\end{enumerate}\reduceitemm
Further, assume that we are able to apply 
$- \reflex ( \emm )$ with cost $c_2$,
and let $\tmax$ be the smallest non-negative integer such that $3^{\tmax} \,
\sin^{-1}\sqrt{\epsilon} \quad \in \quad \left[{\pi}/{4}, {3\pi}/{4}
\right]$, where
$\pM \geq \epsilon > 0$ whenever $\pM>0$.
Then, for every real~$\gamma$ such that 
$0< \gamma\leq \frac{1}{40}$, 
$\textbf{Tolerant RAA}(\tmax,\gamma)$ always
outputs ``No marked element''
if $M$ is empty, 
otherwise it ends with a marked element with probability at least
$1/12-3\gamma$, and incurs a cost of order
$3^{\tmax}\cdot (c_1 \log\tfrac{1}{\gamma} + c_2)$.
\end{lemma}

\begin{proof}
\suppress{
First, if $\emm$ is empty then clearly the algorithm always outputs ``No marked element". 
We now assume that $\emm$ is non-empty and $\pM\geq\eps$.
}
First, if $M$ is empty then clearly the algorithm always outputs
``No marked element". 
We now assume that $M$ is non-empty and $\pM\geq\eps$.
If~$\pM \geq 1/2$, the first two steps of the algorithm succeed with
probability at least~$1/2$. So in the analysis of the remaining steps, 
we additionally assume that~$\pM < 1/2$.

We will use the notations of Lemma~\ref{lemma-grover-approximate-reflection}, together with the following ones. 
For~$i \geq 1$, define $\ket{\psi_i}$ as the state after 
Step~\ref{step:new-attempt}, 
$\sin^2\theta_{i}=\norm{\Pi_\emmt\ket{\psi_{i}}}^2$ the probability to project $\ket{\psi_{i}}$ onto the marked subspace $\emm$,
and the normalized projections $\ket{\nu_{i}}=\Pi_\emmt\ket{\psi_{i}}/\sin\theta_{i}$ and $\ket{\nu_{i}^\perp}=\Pi_{\emmt^\perp}\ket{\psi_{i}}/\cos\theta_{i}$, where $\emmt^\perp$ is the orthogonal complement of $\emmt$.
Initially, we set $\ket{\nu_0^\perp}=\ket{\phi_0}=\ket{\pi}\ket{0^S}$.

Let us denote by $A_i$ the unitary operator corresponding to circuit $\textbf{Approximate RAA}(i,\gamma)$, and let~$t$ be the smallest
positive
integer such that $3^t \,
\sin^{-1}\sqrt{\pM} \quad \in \quad \left[{\pi}/{4}, {3\pi}/{4}
\right]$.
By Lemma~\ref{lemma-grover-approximate-reflection}, Applying $A_t$ on $\ket{\phi_0}=\ket{\pi}\ket{0^S}$ prepares a state $\ket{\phi_t}$ that has projection at least $1/\sqrt{2}-\gamma$ on $\emm$.

Since we do not know $t$, we will apply $A_i$ for all possible values $i\in\left[1,\tmax \right]$. 
To avoid having to prepare a fresh copy of $\ket{\phi_0}$ for each attempt, which would incur an additional cost, for $i>1$ we apply $A_i$ on the state $\ket{\nu_{i-1}^\perp}$ left over from the previous attempt, which produces the state $\ket{\psi_i}=A_i\ket{\nu_{i-1}^\perp}$ instead of $\ket{\phi_i}=A_i\ket{\phi_0}$.

\textbf{Analysis of the errors ---}
Let $\delta_i=\norm{\ket{\psi_i}-\ket{\phi_i}}=\norm{\ket{\nu_{i-1}^\perp}-\ket{\phi_0}}$ denote the error at step $i$. By construction, we have $\delta_1=0$ and, for $i\geq 1$,
\begin{equation}
 \delta_{i+1}=\norm{\ket{\nu_{i}^\perp}-\ket{\phi_0}}
\leq \norm{\ket{\nu_{i}^\perp}-\ket{\mu_{i}^\perp}}
+\sum_{k=0}^{i-1}\norm{\ket{\mu_{k+1}^\perp}-\ket{\mu_{k}^\perp}}
+\norm{\ket{\mu_{0}^\perp}-\ket{\phi_0}}\label{eq:delta_iplus1}.
\end{equation}

Let us evaluate the first term. 
By definition we have
\begin{align}
\ket{\psi_{i}}
&=\sin\theta_{i}\ket{\nu_{i}}+\cos\theta_{i}\ket{\nu_{i}^\perp}\label{eq:psi1},\\
\ket{\phi_{i}}
&=\sin\varphi_{i}\ket{\mu_{i}}+\cos\varphi_{i}\ket{\mu_{i}^\perp}.\nonumber
\end{align}
Since $\norm{\ket{\psi_i}-\ket{\phi_i}}=\delta_i$ we also have
\begin{align}
 \ket{\psi_{i}}&=\sin\varphi_{i}\ket{\mu_{i}}+\cos\varphi_{i}\ket{\mu_{i}^\perp}+\ket{\xi_i}\label{eq:psi2},
\end{align}
where $\norm{\xi_i}\leq \delta_i$. Projecting Equations~(\ref{eq:psi1}) and (\ref{eq:psi2}) onto $\emmt^\perp$, we obtain
\begin{align*}
 \cos\theta_{i}\ket{\nu_{i}^\perp}=\cos\varphi_{i}\ket{\mu_{i}^\perp}+\Pi_{\emmt^\perp}\ket{\xi_i},
\end{align*}
which implies that $|\cos\theta_{i}-\cos\varphi_{i}|\leq\delta_i$ and in turn
\begin{align*}
 \norm{\ket{\nu_{i}^\perp}-\ket{\mu_{i}^\perp}}\leq\frac{2\delta_i}{\cos\varphi_{i}}\leq 3\delta_i,
\end{align*}
for any $i<t$. For the last inequality, we have used the fact that $\bar{\varphi}_{i}<\frac{\pi}{4}$, and therefore $\cos\varphi_{i}\geq\cos\bar{\varphi}_{i}-e_{i}\geq\frac{\sqrt{2}}{2}-\gamma\geq\frac{2}{3}$, 
since $\gamma\leq\frac{1}{40}$.

Let us now evaluate the second term in Equation~(\ref{eq:delta_iplus1}). Recall that
\begin{align*}
 \ket{\phi_{k+1}} &=  \sin\varphi_{k+1} \; \ket{\mu_{k+1}} + \cos\varphi_{k+1} \; \ket{\mu_{k+1}^\perp}\\
&= \sin3\varphi_k \; \ket{\mu_k} + \cos3\varphi_k \; \ket{\mu_k^\perp}+\ket{\omega_{k+1}},
\end{align*}
where
$\norm{\omega_{k+1}} \leq \beta_{k+1} \sin 2\bar{\varphi}_k
\leq 4\beta_{k+1}\bar{\varphi}_k$, by the calculations leading to
Eq.~(\ref{eqn-omega}), and the bound~$e_k \leq \tilde{e}_k \leq 
\bar{\varphi}_k$.
Projecting this equation onto $\emmt^\perp$, we obtain
$|\cos3\varphi_k-\cos\varphi_{k+1}|\leq\norm{\omega_{k+1}}$ and in turn
\begin{align*}
 \norm{\ket{\mu_{k+1}^\perp}-\ket{\mu_{k}^\perp}}
\leq \frac{2\norm{\omega_{k+1}}}{\cos\varphi_{k+1}}
\leq 12\beta_{k+1}\bar{\varphi}_k=12\beta_{k+1}3^k\varphi_0,
\end{align*}
for any $k<t-1$, where we have used the fact that $\cos\varphi_{k+1}\geq\frac{2}{3}$.

For the last term of Equation~(\ref{eq:delta_iplus1}), since $\braket{\mu_0^\perp}{\phi_0}=\cos\varphi_0$, we have
\begin{align*}
 \norm{\ket{\mu_0^\perp}-\ket{\phi_0}}=\sqrt{2-2\cos\varphi_0}=2\sin(\varphi_0/2)\leq \varphi_0.
\end{align*}
Putting everything back together, we have
\begin{align*}
 \delta_{i+1}\leq 3\delta_i+12\varphi_0\sum_{k=0}^{i-1}3^k\beta_{k+1}+\varphi_0,
\end{align*}
which, from $\delta_1=0$, implies
\begin{align*}
 \delta_t&\leq\varphi_0\sum_{k=0}^{t-2} 3^k+12\varphi_0\sum_{k=0}^{t-2}\left(\sum_{j=0}^{t-k-2}3^j\right)3^k\beta_{k+1}\\
&\leq\frac{1}{2}\ 3^{t-1}\varphi_0+12\varphi_0\sum_{k=0}^{t-2}\left(\frac{1}{2}\ 3^{t-k-1}\right)3^k\beta_{k+1}\\
&\leq \frac{1}{2}\ 3^{t-1}\varphi_0+2\cdot 3^t\varphi_0\sum_{k=0}^{t-2}\beta_{k+1}\\
&\leq \frac{\pi}{8}+\frac{9\gamma}{8}.
\end{align*}
Since the projection of $\ket{\phi_t}$ onto $\emmt$ has length at least 
$\frac{1}{\sqrt{2}}-\gamma$, the projection of $\ket{\psi_t}$ onto $\emmt$ has
length at least
$ \frac{1}{\sqrt{2}}-\gamma-\delta_{t}\geq \frac{1}{\sqrt{12}}-3\gamma$, which means that the next measurement projects this state onto $\emmt$ with probability at least $\frac{1}{12}-3\gamma$.

\textbf{Complexity ---} As for the complexity analysis, note that we apply $A_i$ for all $i\in[1,\tmax]$. From Lemma~\ref{lemma-grover-approximate-reflection}, the cost of $A_i$ is
of order
$\mathsf{Cost}(i) \in \Order(3^i \cdot (c_1 \log\tfrac{1}{\gamma} + c_2))$,
therefore the cost of $\textbf{Tolerant RAA}(\tmax,\gamma)$ is
dominated by
\begin{align}
\sum_{i=1}^\tmax\mathsf{Cost}(i)
& \in \Order(3^\tmax \cdot (c_1 \log\tfrac{1}{\gamma} + c_2)),
\end{align}
since this defines a geometric sum.
\end{proof}

We now have all the elements to prove Theorem~\ref{thm-without-log-factor} (stated in
Section~\ref{sec-contribution}).

\begin{proof}[Proof of Theorem~\ref{thm-without-log-factor}]
The algorithm consists in $\textbf{Tolerant RAA}(\tmax,\tfrac{1}{72})$ from 
Lemma~\ref{lemma-unknown-pm}, 
using for the approximate reflections $R(\beta)$
the quantum phase estimation circuit $R(P)$ from Theorem~\ref{thm-diffusion1}.

First, no marked element is found if $M$ is empty.
Assume for now that $M$ is non empty.
We will prove that the assumptions of Lemma~\ref{lemma-unknown-pm}
are satisfied.
Therefore the probability of finding an element for  $\textbf{Tolerant RAA}(\tmax,\tfrac{1}{72})$ is at least $1/24$.
 
Setting
$k= \left\lceil{\log_2(\tfrac{1}{\beta})+ 1}\right\rceil$
in Theorem~\ref{thm-diffusion1}, $R(P)$ simulates a
reflection with an error upper bounded by
$2^{1-k} \leq \beta$.
Implementing $R(P)_d$ then requires $k\,2^{s+1}$ calls to the
controlled quantum walk $\control{W(P)_d}$ or its inverse,
where~$s \in \log_2(\frac{1}{\sqrt{\delta}})+\Order(1)$.
Since implementing
$\control{W(P)_d}$ or its inverse has a cost $4 \mathsf{U}$, the cost
of implementing the circuit $R(P)_d$ for a given error $\beta$ is $c_1
\log\tfrac{1}{\beta}$, with $c_1$ of
order~$\tfrac{1}{\sqrt{\delta}}\mathsf{U}$.  Furthermore, preparing
the initial state $\ket{\pi}_d$ has a cost $\mathsf{S} + \mathsf{U}$,
and implementing $- \reflex (\emm)_d$ has a cost
$c_2=\mathsf{C}$. Finally, 
since~$\tmax \in \log_3 \frac{1}{\sqrt{\eps}} + \Order(1)$,
the total cost of $\textbf{Tolerant RAA}(\tmax,\tfrac{1}{72})$ is of
order $ \mathsf{S} + \frac{1}{\sqrt{\eps}} ( \frac{1}{\sqrt{\delta}}
\mathsf{U}+\mathsf{C}).  $
\end{proof}

\section{Non-reversible Markov chains}
\label{sec-diffusion2}

In this section, we discuss the performance of the search algorithm
presented earlier for
any ergodic, but possibly non-reversible Markov
chain~$P$. 
For the analysis of the quantum walk~$W(P)$ we directly
examine the singular value decomposition of the discriminant
matrix~$D(P) = \diag(\pi)^{1/2} \cdot P \cdot \diag(\pi)^{-1/2}$.
This matrix has the same eigenvalues as~$P$, 
but the singular values of~$D(P)$ may be different from
the eigenvalues of $P$.
The singular values of $D(P)$
lie in the interval~$[0,1]$. The vector~$v = ( \sqrt{\pi_x}\, )$
is both a left and a right eigenvector of~$D(P)$ with
eigenvalue~$1$. Therefore, $\Span(v)$ and~$\Span(v)^\perp$ are
invariant subspaces of~$D(P)$, and we may choose~$v$ to be a left and
right singular vector.
If every singular vector orthogonal to~$v$ has a singular value strictly
smaller than~$1$, that is $D(P)$ has a non-zero singular value gap, then
Theorem~\ref{thm-without-log-factor} and its proof stay valid 
when the eigenvalue gap of $P$ is replaced
by the singular value gap of $D(P)$.

The discriminant of an
irreducible walk does not necessarily have non-zero
singular value gap, even if it is ergodic.
Ergodicity implies a non-zero eigenvalue gap for $P$, but there are
examples of ergodic Markov chains whose
discriminants
have $0$ singular value gap.  In the next proposition
we show that if every state in the Markov chain has a transition to
itself with non-zero probability, then its discriminant has non-zero
singular value gap 
(the proof is given in the appendix).
There is a standard and simple modification to any
Markov chain~$P$ such that the resulting chain has this property: with
some probability ~$\alpha \in (0,1)$, stay at the current state, and
with probability~$1-\alpha$, make a transition according to~$P$.

\begin{proposition}
\label{thm-sv-gap}
Let~$P = (p_{xy})$ be an irreducible Markov chain on a finite state
space~$X$, such that
$p_{xx} > 0$, for every~$x \in X$.    
Then, the discriminant matrix~$D(P)$ 
has exactly one singular value equal to~$1$.
\end{proposition}

Finally, we state the theorem on the performance of the quantum search 
algorithm presented in Section~\ref{sec-search-approx} when
the underlying Markov chain is not necessarily reversible.

\begin{theorem}
\label{thm-svg}
Let~$P = (p_{xy})$ be an irreducible Markov chain on a finite state
space~$X$, such that $D(P)$
has exactly one singular value equal to~$1$.
Let~$\delta > 0$ be the singular value gap of $D(P)$, and let $\eps >
0$ be a lower bound on the probability that an element chosen from the
stationary distribution of $P$ is marked whenever~$M$ is
non-empty. Then, there is a quantum algorithm that with
high probability, determines if $M$ is empty or finds an element of $M$, with cost
of order $ \mathsf{S} + \frac{1}{\sqrt{\eps}} (
\frac{1}{\sqrt{\delta}} \mathsf{U}+\mathsf{C}).  $
\end{theorem}

\section{Acknowledgments}

A part of this work was done while the authors were visiting Institut
Henri Poincar{\'e}, Paris, France, during the Programme on Quantum
Information, Computation, and Complexity, January--April
2006.

This research was partially supported by the European Commission IST 
projects QAP 015848 and QCS 25596, and by the French ANR projects AlgoQP 
and QRAC 08-EMER-012.
A.\ N.\ was supported in part by NSERC Canada, CIFAR, an ERA (Ontario),
QuantumWorks, CFI, OIT, MITACS, and ARO/NSA (USA). 
Research at Perimeter Institute for
Theoretical Physics is supported in part by the Government of Canada
through Industry Canada and by the Province of Ontario through MRI.
Research at the Centre for Quantum Technologies is funded by the
Singapore Ministry of Education and the National Research Foundation.
J.\ R.\ acknowledges support from the Belgian FNRS, NSF Grant CCF-0524837 and ARO Grant
DAAD 19-03-1-0082, and during this work he was affiliated with LRI, Universit\'e Paris-Sud; QuIC, Universit\'e Libre de Bruxelles; and Computer Science Division, U.C. Berkeley.

We thank the anonymous referees for their careful reading of the earlier
drafts of this article, and for suggestions that vastly improved the quality
of presentation.

\bibliography{walk-search}
\bibliographystyle{plain-initials}

\appendix

\section*{Proof of Proposition~\ref{thm-sv-gap}}
We first state and prove that all the singular values of
$D(P)$ lie in $[0,1]$.
\begin{lemma}
\label{thm-sv}
Let~$P = (p_{xy})_{x,y \in X}$ be an irreducible Markov chain with
stationary distribution~$\pi = (\pi_x)_{x \in X}$. Then the singular
values of the matrix~$D(P)$ given by
\[
D(P) \quad = \quad \diag(\pi)^{1/2} \cdot P \cdot \diag(\pi)^{-1/2}
\]
all lie in the interval~$[0,1]$.
\end{lemma}
\begin{proof} 
Singular values are by convention taken to be non-negative real.  To
verify that~$\norm{D(P)}$, the largest singular value of~$D(P)$ is at
most~$1$, consider the inner product~$u^\adjoint D(P) v$, for some
unit vectors~$u,v$. The maximum absolute value that this inner product achieves is
the norm of~$D(P)$. By the Cauchy-Schwarz inequality, the inner
product may be bounded as
\begin{eqnarray}
\nonumber
\lefteqn{\size{u^\adjoint D(P) v} }\\
\nonumber
    & = & \size{\sum_{xy} \bar{u}_x v_y 
          \sqrt{\frac{\pi_x}{\pi_y}} \; p_{xy}} \\
\label{eqn-cs}
    & \leq & \left( \sum_{xy} \size{u_x}^2 p_{xy} \right)^{1/2}
             \left( \sum_{xy} \size{v_y}^2 \frac{\pi_x}{\pi_y}\,
                       p_{xy} \right)^{1/2} \\
\nonumber
    & \leq & 1,
\end{eqnarray}
since~$\sum_x \pi_x p_{xy} = \pi_y$.
\end{proof}

\begin{proof}[Proof of Proposition~\ref{thm-sv-gap}]
{From} Lemma~\ref{thm-sv}, we know that the singular values
of~$D({P})$ all lie in~$[0,1]$. Further~$v = ( \sqrt{\pi_x}\, )$
is a left (and right) singular vector with singular value~$1$. We 
show below that for any left and right singular vectors~$u,w \in
\complex^X$, if~$u^\adjoint \, D({P}) \, w = 1$, then~$u = w =
v$ (modulo an overall phase). This establishes the uniqueness of
the singular value~$1$ and a non-zero singular value gap in~$D(P)$.

Suppose~$u^\adjoint \, D({P}) \, w = 1$. This implies that the
Cauchy-Schwarz inequality in Equation~(\ref{eqn-cs}) in the proof of
Lemma~\ref{thm-sv} is tight. Then necessarily, the two unit
vectors~$u',w' \in \complex^{X \times X}$ given by~$u' = ( u_x
\sqrt{p_{xy}} )_{x,y \in X}$ and\linebreak $w' = (w_y \sqrt{\pi_x p_{xy}/ \pi_y}
)_{x,y \in X}$ are 
parallel. Ignoring an overall phase, we may assume that they are in
fact equal.
This means that for every pair~$x,y \in X$ such that~$p_{xy} > 0$,
$u_x = w_y \sqrt{\pi_x/\pi_y}$.
In particular, since~$p_{xx} > 0$, $u_x = w_x$ for
every~$x$, and so~$u_x = u_y \sqrt{\pi_x/\pi_y}$ for every
neighbor~$y$ of~$x$ in the graph underlying the Markov chain~$P$.

Furthermore, for any path~$x_1, x_2, \ldots, x_k$ in the graph,
chaining together the equations
\begin{eqnarray*}
u_{x_{i+1}} & = & u_{x_i} \sqrt{ \frac{\pi_{x_{i+1}}}{\pi_{x_i}}}, 
\end{eqnarray*}
for~$i = 1, \ldots, k-1$, we get that
\begin{eqnarray*}
u_{x_{i}} & = & u_{x_1} \sqrt{ \frac{\pi_{x_{i}}}{\pi_{x_1}}}, 
\end{eqnarray*}
for every~$i$. Since the chain~$P$ is irreducible, i.e., the
underlying graph is strongly connected, there is a path from~$x_1$
to~$y$ for every~$y \in X$. Thus,
\begin{eqnarray*}
u_{y} & = & u_{x_1} \sqrt{ \frac{\pi_{y}}{\pi_{x_1}}}, 
\end{eqnarray*}
for every~$y$. Since the vector~$u$ is a unit vector, this implies
that~$u = w = ( \sqrt{\pi_x} )_{x \in X} = v$ (up to an unimportant
global phase).
\end{proof}
\end{document}